\documentclass[aps,prd,reprint,superscriptaddress,nofootinbib, longbibliography]{revtex4-2}

\usepackage{graphicx}
\usepackage{dcolumn}
\usepackage{bm}
\usepackage{amsmath,amssymb} 
\usepackage{amsfonts}
\usepackage{subfigure}
\usepackage{graphicx}
\usepackage{ulem}
\usepackage{color}
\usepackage{xcolor}
\usepackage{float}
\usepackage{makecell}
\usepackage{newtxtext,newtxmath}
\usepackage{mathrsfs}
\usepackage{gensymb}
\usepackage{multirow}
\usepackage{booktabs}
\usepackage{threeparttable}
\usepackage{ulem}
\usepackage{rotating}
\usepackage{graphicx}
\usepackage{tabularx}
\usepackage{aas_macros}
\usepackage{verbatim}
\usepackage{url }
\usepackage{caption}
\usepackage{subcaption}
\graphicspath{{figure/}}
\usepackage{hyperref}
\usepackage[mathlines]{lineno}

\newcommand{\dif}{{\rm{d}}}

\DeclareUnicodeCharacter{2212}{-}

\begin{document}

\preprint{APS/123-QED}

\title{Sub-parsec precision measurement of pulsar distances with nanohertz gravitational waves}


\author{Jiming Yu}
\email{jimingyu@sjtu.edu.cn}
\affiliation{Department of Astronomy, School of Physics and Astronomy, Shanghai Jiao Tong University, Shanghai, 200240, China}
\affiliation{Key Laboratory for Particle Astrophysics and Cosmology (MOE) / Shanghai Key Laboratory for Particle Physics and Cosmology, China}
\author{Zhen Pan}
\email{zhpan@sjtu.edu.cn}
\affiliation{Tsung-Dao Lee Institute, Shanghai Jiao-Tong University, Shanghai, 520 Shengrong Road, 201210, China}
\affiliation{Department of Astronomy, School of Physics and Astronomy, Shanghai Jiao Tong University, Shanghai, 200240, China}

\date{\today}

\begin{abstract}
The recent evidence of nanohertz (nHz) gravitational wave (GW) background by pulsar timing array (PTA) collaborations has sparked considerable interest in understanding its astrophysical origins, particularly regarding supermassive black hole binaries (SMBHBs). 
In this work, we focus on  individual SMBHBs that will be hopefully detected 
in upcoming PTA observations. The effect of nHz GWs on the pulse arriving times is in general decomposed as a pulsar term and an Earth term, where
the pulsar term encodes the pulsar-Earth distance as a phase shift relative to the Earth term, 
but is usually treated as an extra noise source since the pulsar distance 
is in general not well measured with uncertainty larger than the wavelength of nHz GWs. We propose that the pulsar distance could be  constrained by combining the phase information of multiple SMBHBs that are individually resolved. Using Markov chain Monte Carlo (MCMC) simulations, we demonstrate that the pulsar distances can be measured to better than $0.4$ pc (1 pc) for pulsars at $D\sim 1$ kpc ($\sim 2.2$ kpc) with 30 years of observations  by a 20-pulsar PTA with a noise level of $\sigma_{\rm n}=20$ ns in the Square Kilometre Array (SKA) era.
\end{abstract}

\maketitle

\section{Introduction} \label{sec:intro}

As gravitational waves (GW) propagate through the Milky Way,  their spacetime perturbations induce detectable variations in the arrival times of pulsar signals \cite{1975GReGr...6..439E}. 
Thereby, millisecond pulsars (MSPs \cite{1990ApJ...361..300F}) have been used as GW detectors to monitor the nanohertz (nHz) GWs in the universe by observing the pulse times of arrival (TOAs) \cite{1978SvA....22...36S, 1979ApJ...234.1100D}, which are named as the pulsar timing array (PTA). Recent breakthroughs by several collaborations, including the European Pulsar Timing Array (EPTA \cite{2008AIPC..983..633J}), the Parkes Pulsar Timing Array (PPTA \cite{2008AIPC..983..584M}), the North American Nanohertz Observatory for Gravitational Waves (NANOGrav \cite{2009arXiv0909.1058J}), the Chinese Pulsar Timing Array (CPTA \cite{2016ASPC..502...19L}), the MeerKAT Pulsar Timing Array (MPTA \cite{2023MNRAS.519.3976M}) have revealed a GW background signal at $2-4.6 \ \sigma$ confidence levels through the correlation between the timing residuals of pulsar pairs \cite{2023ApJ...951L...8A, 2023A&A...678A..50E, 2023ApJ...951L...6R, 2023RAA....23g5024X, 2025MNRAS.536.1467M}, which is the well known  Hellings and Downs (HD) curve \cite{1983ApJ...265L..39H}. 

Supermassive  black hole binaries (SMBHBs) with sub-parsec binary separations are natural sources of  nHz stochastic GWs. These systems originate from galactic mergers, where two SMBHs are brought into a same galaxy and form a  bound binary through dynamical processes \cite{1980Natur.287..307B}. During their hierarchical evolution, SMBHBs lose energy via stellar scattering and viscous gas interactions  \cite{2015ApJ...810..139H, 2017MNRAS.464.2301G, 2018MNRAS.473.3410R, 2019MNRAS.486.4044B}, until they transition decisively into the GW-driven regime \cite{2003AIPC..686..201M}. 
There are also plenty of models attributing the nHz stochastic GWs to various early Universe processes.  To distinguish the two origins, efforts are made in searching for either signatures of individual SMBHBs \cite{2023ApJ...951L..50A, 2024A&A...690A.118E,2025MNRAS.536.1501G}, or non-Gaussian fluctuations in the signals expected from a collection of SMBHBs \cite{Ellis:2023owy,Sato-Polito:2024lew,Xue:2024qtx}.
Recently, the MPTA collaboration reported a hotspot in 7 nHz GW sky map with a $p$ value of 0.015  \cite{2025MNRAS.536.1501G}, which may have an astrophysical origin (a SMBHB), although an origin of noise fluctuations cannot be confidently ruled out yet.  
Though none of the efforts above lead to a confident detection of individual SMBHBs  limited by the current detection accuracy and observation duration, many works predict that they will hopefully be detected in the near future \cite{2008MNRAS.390..192S, 2009MNRAS.394.2255S, 2020ApJ...897...86C, 2020PhRvD.102b3014F, 2024arXiv240701659C, 2024ApJ...974..261C}.

When a radio pulse/a photon propagates from a pulsar to the Earth, its TOA (or equivalently frequency) will be perturbed by the stochastic GWs. The net effect 
is integrated as the difference between two terms that are proportional to the GW strength at the pulsar $\boldsymbol{x}_{\rm p}$ when the pulse is emitted $t_{\rm p}$
and at the Earth $\boldsymbol{x}_{\rm E}$ when the pulse is received $t_{\rm E}$, respectively, and are therefore called the pulsar term  and the Earth term  \cite{1975GReGr...6..439E, 1979ApJ...234.1100D, 1987GReGr..19.1101W, 2010PhRvD..81j4008S}. 
For typical pulsars used in PTAs, the timescale $|t_{\rm E}-t_{\rm p}|$ is of $\mathcal{O}(10^3)$ years, the frequencies of the two terms will differ due to the binary evolution, which is imprinted in the waveform of the timing residuals \cite{2010arXiv1008.1782C}. Since the phase difference  between the two terms is proportional to the pulsar distance $D$, it is natural to infer $D$ from the timing residuals. The biggest challenge in this application is that the perturbation caused by the nHz GW has a period of about 1 year, which results in a period of $\sim1$ ly in the constraint of pulsar distance $D$.
To solve this problem, Lee et al. (2011) \cite{2011MNRAS.414.3251L} proposed that the high-precision timing parallax measurements of the millisecond pulsars \cite{1986ARA&A..24..537B, 1991ApJ...371..739R, 2004hpa..book.....L, 2010MNRAS.405..564V} could be utilized to eliminate the periodicity in the pulsar distance constraint, with the help of the high precision measurements from future radio telescopes, like the Square Kilometre Array (SKA) \cite{2011A&A...528A.108S}. 

Generally, whether the periodicity in the pulsar distance $D$ constraint can be eliminated is determined by three main factors: the wavelength of the GW $\lambda$, the angle between the source and the pulsar $\theta$, 
the uncertainty of the timing parallax measurements $\sigma_D$. The first two factors determine the length period in the direction of pulse propagation perturbed by GWs, and hence determine the period  in the $D$ constraint, therefore, a larger GW wavelength and a smaller $\theta$ correspond to a lager period in the $D$ constraint. And the ratio between the timing parallax measurement uncertainty $\sigma_D$ and this period determines the ability of suppressing those `false' peaks in the $D$ constraint.    Based on \cite{2011MNRAS.414.3251L}, in the first half of this paper, we use Markov chain Monte Carlo (MCMC) simulations and give a more detailed and quantitative analysis of the feasibility of the timing parallax measurements in eliminating the periodicity. 

As we will show later, for pulsars with $D\gtrsim1$ kpc, the periodicity in the $D$ constraint inferred from PTA observations of each individual SMBHB cannot be eliminated due to the limited accuracy of timing parallax measurements. On the other hand, these constraints from GW signals of different SMBHBs  usually have different periods due to the their different wavelengths and sky locations. Based on this fact, we propose that if these distance constraints are combined, 
the false peaks in the $D$ constraint can be eliminated.  

Our paper is organized as follows. In Sec. \ref{sec:waveform}, we introduce the formula for the timing residuals and the PTA's responses. In Sec. \ref{sec:methodology}, we introduce the MCMC method and how it is used to obtain the pulsar distance posteriors. We show the constraints from a single SMBHB and multiple binaries in Sec. \ref{sec:results}. And  we conclude this paper with Sec. \ref{sec:conclusions}. 
Throughout the paper, we use the geometric units with $G=c=1$, and 
assume the standard $\Lambda$CDM model with parameters from the latest Planck results, $\Omega_\mathrm{m}=0.3089$, $\Omega_\mathrm{b}=0.0486$, $\Omega_{\Lambda}=0.6911$, $h=0.6774$ 
And we adopt 68\% confidence intervals (CI) when showing the uncertainties of model parameter constraints  if not  specified otherwise.

\section{PTA observable}
\label{sec:waveform}

\subsection{Timing Residuals}

For a GW source with right ascension (RA)  $\alpha$ and declination (DEC) $\delta$, the GW tensor at $(t, \boldsymbol{x})$ can be written as
\begin{equation}
    h_{ij}(t-\boldsymbol{x}\cdot\hat{\Omega})=\epsilon_{ij}^{+}h_+(t-\boldsymbol{x}\cdot\hat{\Omega})+ \epsilon_{ij}^{\times}h_\times(t-\boldsymbol{x}\cdot\hat\Omega),
\end{equation}
where $\hat{\Omega}=-(\cos\delta\cos\alpha,\ \cos\delta\sin\alpha,\ \sin\delta)$ is the GW propagation direction. The two polarization tensors $\epsilon_{ij}^A(\hat{\Omega})\ (A=+,\ \times)$ above are defined by
\begin{equation}
    \epsilon_{ij}^{+}\equiv\hat{\alpha}_i\hat{\alpha}_j-\hat{\delta}_i\hat{\delta}_j,
\end{equation}
\begin{equation}
    \epsilon_{ij}^{\times}\equiv\hat{\alpha}_i\hat{\delta}_j+\hat{\alpha}_j\hat{\delta}_i,
\end{equation}
where $\hat{\alpha}$ and $\hat{\delta}$ are two unit vectors orthogonal to $\hat{\Omega}$,
\begin{equation}
    \hat{\alpha}=(-\sin\alpha,\ \cos\alpha,\ 0),
\label{eq_alpha}
\end{equation}
\begin{equation}
    \hat{\delta}=(-\sin\delta\cos\alpha,\ -\sin\delta\sin\alpha,\ \cos\delta).
\label{eq_delta}
\end{equation}
Consider a pulsar that emits pulses of frequency of $\nu_0$ along the direction $\hat{p}$. As the pulses propagate, they will be perturbed by $h_{ij}(t-\boldsymbol{x}\cdot\hat{\Omega})$ and result in the following changes in the frequency of the pulses received by the observer \cite{1975GReGr...6..439E, 1979ApJ...234.1100D, 1987GReGr..19.1101W, 2010PhRvD..81j4008S,  Niu:2018oox}, 
\begin{equation}
\begin{aligned}
    z(t,\hat{\Omega})&\equiv\frac{\nu(t)-\nu_0}{\nu_0}\\
    &=\frac{1}{2}\sum_{A=+,\times}\left[ F^A\Delta h_A(t,\hat{\Omega})\right],
\label{eq_frequency}
\end{aligned}
\end{equation}
where $F_A$ are the antenna beam patterns, which are defined as
\begin{equation}
    F^{A}\equiv\frac{\hat{p}_i\hat{p}_j}{1+\hat{\Omega}\cdot\hat{p}}\epsilon^A_{ij}(\hat{\Omega}),\quad A=+,\times,
\end{equation}
and the term $\Delta h_A(t, \hat{\Omega})$ represents the difference of the GW tensor between the pulsar and the observer.
For Earth observers, $\Delta h_A(t,\Omega)$ can be written as
\begin{equation}
    \Delta h_{A}(t,\Omega)=h_{A}^\mathrm{P}-h_{A}^\mathrm{E}=h_{A}(t-D-D\hat{\Omega}\cdot\hat{p})-h_{A}(t),
\label{eq:Dh}
\end{equation}
here $D$ is the pulsar distance, $h_{A}^\mathrm{P}$ and $h_{A}^\mathrm{E}$ are the pulsar term and Earth term, respectively. In practice, time residuals are used as the PTA observable, which  are  the integrals of $z(t,\hat{\Omega})$ from the start of the observation,
\begin{equation}
    R(t)=\int_{0}^{t}\dif t'\ z(t',\hat{\Omega}).
\end{equation}

\subsection{GW waveform}
In this work, we focus on GW signals from circular SMBHBs. In the plane wave approximation, the waveform can be written as \citep{Maggiore:2007ulw}
\begin{equation}
\begin{aligned}
    h_+(t)=h_0[&\cos\iota\sin2\psi\sin(\omega t+\Phi_0)\\
    &-\frac{1}{2}(1+\cos^2\iota)\cos2\psi\cos(\omega t + \Phi_0)],
    \label{eq_h1}
\end{aligned}
\end{equation}
\begin{equation}
\begin{aligned}
    h_\times(t)=-h_0[&\cos\iota\cos2\psi\sin(\omega t + \Phi_0)\\
    &+\frac{1}{2}(1+\cos^2\iota)\sin2\psi\cos(\omega t + \Phi_0)],
    \label{eq_h2}
\end{aligned}
\end{equation}
where $\iota$ is the inclination angle, $\psi$ is the polarization angle, $\omega$ is the GW angular frequency, $\Phi_0$ is the initial phase, and the GW amplitude $h_0$ is
\begin{equation}
    h_0=\frac{2^{8/3}\mathcal{M}_c^{5/3}}{d_L}\omega^{2/3},
\label{eq_h02}
\end{equation}
where 
\begin{equation}
    \mathcal{M}_c\equiv \frac{(m_1 m_2)^{3/5}}{(m_1+m_2)^{1/5}}(1+z)
\end{equation}
is the redshift chirp mass of the SMBHB, $d_L$ is the luminosity distance, and $z$ is the redshift. 
The power of  GW radiation from a circular binary is known as  \cite{Maggiore:2007ulw}
\begin{equation}
    P=\frac{2^{5/4}(\mathcal{M}_c\omega)^{10/3}}{5},
\end{equation}
and the total energy of the binary is
\begin{equation}
    E=-2^{-5/3}\mathcal{M}_c^{5/3}\omega^{2/3},
\end{equation}
therefore, the rate of $\omega$ change over time is 
\begin{equation}
    \dot{\omega}=\frac{3}{5}2^{7/3}\mathcal{M}_c^{5/3}\omega^{11/3}.
    \label{eq_domega}
\end{equation}
Usually, during the observation period of the PTA, the angular frequency change during the observation period $\dot{\omega}T\ll\omega$ is negligible. However, for the pulsar term $h_{ij}^\mathrm{P}$, its time difference from the Earth term is 
\begin{equation}\label{eq:Delta_t}
    \Delta t \equiv D(1+\hat{\Omega}\cdot \hat{p})=D(1-\cos\theta)\gg T,
\end{equation}
where $\cos\theta\equiv-\hat{\Omega}\cdot \hat{p}$, and the difference in the angular frequencies of the two terms is $\Delta\omega \approx \dot{\omega}\Delta t$.
For $\mathcal{M}_c=10^{10}\ M_\odot, \Delta t =10^3$ years, $\omega=30$ nHz event, $\Delta\omega\approx1.6$ nHz. We assume the the angular frequencies of the pulsar and Earth terms $\omega_\mathrm{P}$ and $\omega_\mathrm{E}$ 
remain constant during the PTA observation period, and adopt this ``bichromatic wave approximation'' in the following  calculations.

Finally, the timing residual can be written as 
\begin{equation}
    \begin{aligned}
        R(t)&=\frac{-(2\mathcal{M}_c)^{5/3}}{d_L}\\
        &\times\left\{\left [ \omega_\mathrm{P}^{-\frac{1}{3}}\cos(\omega_\mathrm{P}t+\Phi_0-\Phi_\mathrm{p})\right.-\omega_\mathrm{E}^{-\frac{1}{3}}\cos(\omega_\mathrm{E}t+\Phi_0)\right ]\\ 
        &\quad\times\cos\iota(F^+\sin2\psi-F^\times\cos2\psi)\\
        &+\left [ \omega_\mathrm{P}^{-\frac{1}{3}}\sin(\omega_\mathrm{P}t+\Phi_0-\Phi_\mathrm{p})-\omega_\mathrm{E}^{-\frac{1}{3}}\sin(\omega_\mathrm{E}t+\Phi_0)\right]\\
        &\quad \left.\times\frac{1+\cos^2\iota}{2}(F^+\cos2\psi-F^\times\sin2\psi)\right\},
    \end{aligned}
    \label{eq_Rt}
\end{equation}
where $\Phi_0$ is the initial phase of the Earth term, $\Phi_\mathrm{P}$ is the phase difference between the Earth term and the pulsar term. In the linear approximation,
\begin{equation}
    \omega_\mathrm{P}\approx\omega_\mathrm{E}-\dot{\omega}_\mathrm{E}\Delta t,
    \label{eq_omegaP}
\end{equation}
and
\begin{equation}
    \Phi_\mathrm{P}\approx\omega_\mathrm{E}\Delta t-\frac{1}{2}\dot{\omega}_\mathrm{E}\Delta t^2.
    \label{eq_PhiP}
\end{equation}
Note that when a binary has a high $\omega$ and is close to merging, the GW frequency evolves so fast that the above approximation breaks down. 
Fortunately, since the rate of SMBHB sources at $\omega$ is proportional to $\dot{\omega}^{-1}$, the linear approximation remains valid for the majority of SMBHBs.
Therefore, we ignore SMBHBs exhibiting rapid frequency evolution and assume all SMBHBs satisfy  Eqs.~ (\ref{eq_omegaP}) and (\ref{eq_PhiP}) in this work.

\subsection{Pulsar Timing Array}
Following the setup in Ref.~\cite{2020PhRvD.102b3014F},  we consider a small SKA PTA with $N_\mathrm{psr}=20$ pulsars and assume a  white noise with standard deviation $\sigma_n=20$ ns in TOAs of each pulsar.
For the pulsar positions,  we select 20 pulsars with timing parallax measurements in the EPTA DR2 \cite{2023A&A...678A..48E} (see Table \ref{tab:pulsars} for more information).  
We also consider three different observation spans with $T=10,\ 20,\ 30$ years, respectively, and an observation cadence with $\Delta T=2$ weeks. 
The number of each pulsar's TOA measurements  $N_\mathrm{obs}$ is  therefore $T/\Delta T$. The signal-to-noise ratio (SNR) of the $i$-th pulsar's timing residuals is given by
\begin{equation}
    \mathrm{SNR}^2_i=\sum_{j=1}^{N_\mathrm{obs}}\left[\frac{R_{i}^2(t_j)}{\sigma_n^2}\right],
\end{equation}
where $R_i$ is the timing residual of the $i$-th pulsar, $t_j$ is the time of the $j$-th measurement, and the total SNR is
\begin{equation}
    \mathrm{SNR}^2=\sum_{i=1}^{N_\mathrm{psr}}\mathrm{SNR}^2_i.
\end{equation}
In this work, we adopt $\mathrm{SNR}=8$ as the detection threshold.

\begin{table}[ht]
    \centering
    \begin{tabular}{cccc}
    \hline\hline
        Name & RA& DEC&$D$ (kpc) \\
        \hline
        J0030+0451 & 0h30m27.4s & 0h19m26.6s & 0.323 \\
        J0613-0200 & 6h13m44.0s & -0h08m03.1s & 0.99 \\
        J0751+1807 & 7h51m09.2s & 1h12m30.6s & 1.17 \\
        J1012+5307 & 10h12m33.4s & 3h32m28.2s & 1.07 \\
        J1022+1001 & 10h22m58.0s & 0h40m07.5s & 0.85 \\
        J1024-0719 & 10h24m38.7s & -0h29m17.3s & 0.98 \\
        J1455-3330 & 14h55m48.0s & -2h14m03.1s & 0.76 \\
        J1600-3053 & 16h00m51.9s & -2h03m35.3s & 1.39 \\
        J1640+2224 & 16h40m16.7s & 1h29m36.6s & 1.08 \\
        J1713+0747 & 17h13m49.5s & 0h31m10.5s & 1.136 \\
        J1730-2304 & 17h30m21.7s & -1h32m18.1s & 0.48 \\
        J1744-1134 & 17h44m29.4s & -0h46m19.6s & 0.388 \\
        J1751-2857 & 17h51m32.7s & -1h55m51.1s & 0.79 \\
        J1801-1417 & 18h01m51.1s & -0h57m10.3s & 1.0 \\
        J1804-2717 & 18h04m21.1s & -1h49m10.1s & 0.8 \\
        J1857+0943 & 18h57m36.4s & 0h38m53.1s & 1.11 \\
        J1909-3744 & 19h09m47.4s & -2h30m57.0s & 1.06 \\
        J1911+1347 & 19h11m55.2s & 0h55m10.3s & 2.2 \\
        J1918-0642 & 19h18m48.0s & -0h26m50.3s & 1.3 \\
        J2124-3358 & 21h24m43.8s & -2h15m55.0s & 0.47 \\
        \hline
    \end{tabular}
    \caption{The pulsars used in this paper, where we select 20 pulsars
with timing parallax measurements in the EPTA DR2 \cite{2023A&A...678A..48E}. More details can be found in \cite{2004IAUS..218..139H}.}
    \label{tab:pulsars}
\end{table}

\section{Methodology}
\label{sec:methodology}
In the above section, we describe the response of the PTA to GWs from SMBHBs. As shown in Eq.~ (\ref{eq_Rt}), the observed timing residuals   $R(t)$ contain information about the pulsar distance $D$ in two aspects: 1) the frequency evolution $\Delta\omega/\dot\omega$ and 2) the phase difference $\Phi_{\rm p}$. Since the observable $R(t)$ is a periodic function of $\Phi_{\rm p}$ with a period  $2\pi$,  we expect a broad (and usually single-peak) constraint on $D$ from the  frequency evolution, and a multi-peak constraint from the phase difference $\Phi_{\rm p}$ due to the period $2\pi$ in $\Phi_{\rm p}$.
In this section, we will introduce how to use Bayes' theorem to extract this information and obtain the constraints of $D$.

\subsection{Bayesian Parameter Estimation}
From the Bayes's theorem, the posterior of parameters $\boldsymbol{\theta}$ can be written as
\begin{equation}
    P(\boldsymbol{\theta}|\boldsymbol{R})\propto \mathcal{\pi}(\boldsymbol{\theta})\mathcal{L}(\boldsymbol{R}|\boldsymbol{\theta}),
\end{equation}
where $\boldsymbol{R}$ is the data, $\mathcal{\pi}(\boldsymbol{\theta})$ is the prior, $\mathcal{L}(\boldsymbol{R}|\boldsymbol{\theta})$ is the likelihood. For the timing residuals, the log-likelihood is
\begin{equation}
    \log\mathcal{L}(\boldsymbol{R}|\boldsymbol{\theta})=-\frac{1}{2}\sum_{i=1}^{N_\mathrm{psr}}\sum_{j=1}^{N_\mathrm{obs}}\left[\frac{R_{i}(\boldsymbol{\theta},t_j)-\mu(\boldsymbol{\theta}_0,t_j)}{\sigma_n}\right]^2,
\end{equation}
where $\mu(\boldsymbol{\theta}_0,t_j)$ represents the timing residuals with injection parameters $\boldsymbol{\theta}_0$. With the above log-likelihood, we use the MCMC method in \texttt{bilby} \cite{2019ApJS..241...27A} to estimate the posterior of $\boldsymbol{\theta}$. 

From Eq.~(\ref{eq_Rt}), the timing residuals contain $8$ source parameters, $(\alpha,\ \delta,\ \log(d_L),\ \iota,\ \mathcal{M}_c,\ \psi,\ \phi_0,\ \omega_\mathrm{E})$, and $2N_\mathrm{psr}$ pulsar parameters, $\omega_{\mathrm{P},i}$ and $\Phi_{\mathrm{P},i}$, with $i=1, 2, 3,\cdots,N_\mathrm{psr}$. It is important to note that the two sets of pulsar parameters are not independent, they can both be obtained from $D_i$ and other source parameters via Eqs.~(\ref{eq_omegaP}) and (\ref{eq_PhiP}). There are actually only $8+N_\mathrm{psr}$ independent parameters in the timing model. However, if $D_i$ are directly treated as parameters, the likelihood function will appear a high level of periodicity about $D_i$,
as explained in the beginning of this Section. As a result, model parameter sampling will suffer severe efficiency problem due to fragmented parameter space exploration.
To solve the parameter sampling efficiency problem caused by the periodicity, we decompose the Bayesian inference as two steps: 
1) we treat $D_i$ and   $\Phi_{\mathrm{P},i}$ as independent parameters,  and set $\Phi_{\mathrm{P},i}$ as a variable in the range of $(0,2\pi]$  in the parameter inference; 
 2) the information of $D_i$ is encoded in both the $D_i$ posterior and the $\Phi_{\mathrm{P},i}$ posterior. We obtain the full constraint on $D_i$ by  post-processing the MCMC samples in the $\Phi_{\mathrm{P},i}$'s parameter space. Details will be introduced in the following subsection.

In  step 1) of the above procedure, we find $D_i$ is generally poorly constrained, 
where $D_i$ is constrained from the frequency evolution $\Delta\omega_i=\omega_{{\rm P}, i}-\omega_{\rm E}$ while $\omega_{\mathrm{P},i}$ is loosely constrained due to $\mathrm{SNR}_i\ll\mathrm{SNR}$.
In comparison, the timing parallax measurement of the pulsar distance $D_i$ is more informative, which we therefore use as a prior. 
As a result, the majority of the $D$ information is encoded in the prior and in the phase $\Phi_{\mathrm{P},i}$.


Another point to note is that when the SMBHB event is localized well, there will be a substantial chance to uniquely identify the host galaxy through multi-messenger observations \cite{2024arXiv240604409P}. 
In this case, we replace the priors of $\alpha,\ \delta,\ \log(d_L)$ with delta functions peaked at their injection values,
and these three parameters will not be randomly sampled in the MCMC simulations \cite{2019ApJS..241...27A}. Considering the accuracy of SKA, in this paper we focus on the discussion of the case of SMBHBs with host galaxies identified.

Table \ref{tab:priors} lists the parameters and priors used in MCMC simulations. As an estimate of  uncertainties of  timing parallax measurements $\sigma_D$, we adopt the formula in \cite{2011MNRAS.414.3251L}
\begin{equation}
    \sigma_D \simeq \frac{2.34}{\cos^2\beta_\mathrm{P}}\left( \frac{N_\mathrm{obs}}{100}\right)^{-\frac{1}{2}}
    \left(\frac{D}{1\ \mathrm{kpc}}\right)^2\left(\frac{\sigma_n}{10\ \mathrm{ns}}\right)\ \mathrm{pc},
\label{eq_sigma_D}
\end{equation}
where $\beta_\mathrm{P}$ is the ecliptic latitude of the pulsar. In Table \ref{tab:sigma_D}, each pulsar's $\sigma_D$ with different observation timing spans is listed. 

\begin{table}[htbp]
    \centering
    \begin{tabular}{cccc}
    \hline\hline
        Parameter & Prior & Minimum & Maximum\\
        \hline
        $\alpha$ & Delta & $\alpha$ &\\
        $\delta$ & Delta & $\delta$ &\\
        $\log(d_L)$ & Delta & $\log(d_L)$ &\\
        $\iota$ & Sine & 0 & $\pi$\\
        $\mathcal{M}_c$ & LogUniform & $0.1\mathcal{M}_c$ & $10\mathcal{M}_c$ \\
        $\psi$ & Uniform & 0 & $\pi$\\
        $\phi_0$ & Uniform & 0 & $2\pi$\\
        $\omega_\mathrm{E}$ & Uniform & 0 & $2\omega_\mathrm{E}$\\
        $D_{i}$ & Gaussian & $\mathcal{N}(D_{\mathrm{par},i}, \sigma_{D,i})$ & \\
        $\Phi_{\mathrm{P},i}$ & Uniform & 0 & $2\pi$\\
        \hline
    \end{tabular}
    \caption{The priors used in simulations. }
    \label{tab:priors}
\end{table}

\begin{table}[ht]
    \centering
    \begin{tabular}{cccc}\hline\hline
        $\sigma_D$(pc)&  10 years & 20 years & 30 years \\\hline
        J0030+0451&0.30&0.21&0.17\\
        J0613-0200&3.48&2.46&2.01\\
        J0751+1807&3.98&2.81&2.29\\
        J1012+5307&5.45&3.86&3.15\\
        J1022+1001&2.09&1.48&1.21\\
        J1024-0719&3.01&2.13&1.74\\
        J1455-3330&1.81&1.28&1.05\\
        J1600-3053&5.77&4.08&3.33\\
        J1640+2224&6.54&4.63&3.78\\
        J1713+0747&5.06&3.58&2.92\\
        J1730-2304&0.67&0.47&0.39\\
        J1744-1134&0.46&0.32&0.26\\
        J1751-2857&1.82&1.29&1.05\\
        J1801-1417&2.97&2.10&1.72\\
        J1804-2717&1.86&1.32&1.08\\
        J1857+0943&5.00&3.53&2.89\\
        J1909-3744&3.49&2.47&2.02\\
        J1911+1347&21.36&15.10&12.33\\
        J1918-0642&5.26&3.72&3.04\\
        J2124-3358&0.71&0.50&0.41\\ \hline
    \end{tabular}
    \caption{The uncertainty $\sigma_D$ of pulsar distance measured with timing parallax with different observation timing spans.}
    \label{tab:sigma_D}
\end{table}

\subsection{Pulsar Distance Constraints}
After obtaining the posterior of $\Phi_{\mathrm{P},i}$ in step 1), step 2) is to convert the constraint on $\Phi_{\mathrm{P},i}$ to a constraint on $D_{i}$. 
From Eq.~(\ref{eq_PhiP}), the posterior of $D_i$ can be obtained from
\begin{equation}
    \mathcal{P}(D_i)\propto \sum_{k}\mathcal{P}[D(\Phi_{\mathrm{P},i}+2k\pi, \omega_\mathrm{E}, \mathcal{M}_c, \cos\theta)],
\label{eq_PDi}
\end{equation}
where $ k=0,1,2,\cdots$.
In general, this posterior consists of peaks with a roughly same shape, and the separation between the $k$-th and $(k+1)$-th peaks is about 
\begin{equation}
     \delta D_{k,k+1}\approx2\pi/(1-\cos\theta)\omega_\mathrm{E}.
    \label{eq_deltaD}
\end{equation}
For the SKA, the width of the $k$-th peak $\Delta D_k$ is mainly determined by the uncertainty of $\omega_\mathrm{E}$,
\begin{equation}
    \Delta D_k/D\approx\Delta\omega_\mathrm{E}/\omega_\mathrm{E}\ .
\end{equation}
If $\Delta D_k \gtrsim \delta D_{k,k+1} / 2$ when $\Delta \omega_\mathrm{E}\gtrsim\pi/\Delta t$,  different peaks of $D_i$'s posterior will have a significant overlap, so that the final constraints become nearly uniformly distributed. This means the pulsar distance will not be able to be constrained through the $\Phi_{\mathrm{P},i}$ measurements. For $\Delta t=10^3$ years, the threshold is about $\Delta \omega_{\mathrm{E,thr}}\equiv\pi/\Delta t\approx 0.2$ nHz. 
In general, a high SNR$\gg8$ PTA observation is required for obtaining a desired constraint with $\Delta \omega_{\mathrm{E}}<\Delta \omega_{\mathrm{E,thr}}$ unless $\cos\theta\sim1$.





In combination with the prior information $\pi(D_i)$ in Table~\ref{tab:sigma_D}, we obtain the full constraint on $D_i$ from PTA observations of a single SMBHB as 
\begin{equation}
    \mathrm{PDF}(D_i)=\pi(D_i)\mathcal{P}(D_i),  
\label{eq:PD1}    
\end{equation}
where  $\pi(D_i):= \mathcal{N}(D_i, \sigma_{D,i})$ is a Gaussian function, and  $\mathcal{P}(D_i)$ is a periodic function defined in Eq.~(\ref{eq_PDi}). The `false' peaks in $\mathcal{P}(D_i)$ will be suppressed by the prior information from timing parallax measurements. 
In the case of $\sigma_{D,i} \ll \delta D_{k,k+1}$, the periodicity in $D_i$'s posterior will be eliminated completely. Meanwhile, if the error bar of a single peak in $\mathcal{P}(D_i)$ is much smaller than $\sigma_{D,i}$, PTA observations of such SMBHB  can significantly improve pulsar distance measurements and we label them as golden events. We take $\delta D_{k, k+1} > 3\sigma_{D,i}$ as the threshold of golden events, which requires
\begin{equation}
    \cos\theta\gtrsim 1 - \frac{2\pi}{3\omega_\mathrm{E}\sigma_{D,i}}.
\label{eq:golden}
\end{equation}
In particular, when
\begin{equation}
    \frac{2\pi}{3\omega_\mathrm{E}\sigma_{D,i}} > 2,
\label{eq:golden2}
\end{equation}
SMBHBs in all directions satisfy Eq.~(\ref{eq:golden}).

The above analysis is suitable for pulsars with high timing parallax measurement accuracy with  $\sigma_{D,i} \lesssim \delta D_{k, k+1}$. 
However, when $\sigma_{D,i} \gtrsim \delta D_{k,k+1}$, the timing parallax measurement is unable to eliminate the periodicity in $D_i$'s posterior. In such case, it is necessary to find another way to eliminate the periodicity of $D_i$.

Generally, the posteriors of $D_i$  from different SMBHBs typically have different periods due to 
different binary frequencies $\omega_\mathrm{E}$ and different sky locations $\cos\theta$, and `fake' peaks in the constraints obtained from observations of different SMBHBs will be located at different values. This property thus can be used to solve the periodicity in $D_i$'s constraint.
Combining all the information form  $N$ different SMBHBs, the full constraint becomes
\begin{equation}
    \mathrm{PDF}(D_i)=\mathcal{\pi}(D_{i})\prod_{n=1}^{N}\mathcal{P}_n(D_i)\ .
    \label{eq:multi}
\end{equation}
The `fake' peaks in $\mathcal{P}_n(D_i)$ will have the potential to cancel each other out, ultimately leaving only a single sharp true peak.  In the next section, we will discuss these two methods of eliminating periodicity.

\section{Results}
\label{sec:results}

\subsection{Constraints from Individual Sources}
Simulations in \cite{2024arXiv240701659C} show that chirp masses $\mathcal{M}_c$ and frequencies of SMBHBs detectable by PTA are mainly distributed around $\mathcal{M}_c\sim10^{10}\ M_\odot$ and $f=3-15$ nHz, respectively. Though their simulations are based on an assumed near-future PTA, the results may vary considerably between different PTAs due to observed selection effects. However, the detectable SMBHBs in their simulations still represent the sources with the highest SNRs, which are crucial for pulsar distance constraints. Based on these simulations, we consider the SMBHB samples with $\mathcal{M}_c=10^{10}\ M_\odot,\ \omega_\mathrm{E}=10,\ 30$ nHz, and observation periods of $T=10,\ 30$ years. For each case of $\mathcal{M}_c,\ \omega_\mathrm{E},\ T$, we sample 20 SMBHBs with an isotropic distribution in teh sky, uniform distributions in $\cos\iota,\ \psi,\ \Phi_0$, and $d_L=$ 5, 20 Gpc, respectively. Here $d_L=5$ Gpc is roughly the lower limit of a $\mathcal{M}_c=10^{10}\ M_\odot$ SMBHB according to recent PTA data releases \cite{2023ApJ...951L..50A, 2023arXiv230616226A}, and $d_L=20$ Gpc is close to the upper limit  that can be used for pulsar distance constraints in the tests. We simulate the observations of each event, conduct the parameter inference using MCMC simulations, and present the pulsar distance measurement results for J0030+0451, J0613-0200, J1911+1347 as examples of pulsars with $D \approx 0.3$, 1, and 2 kpc, respectively.

\begin{figure}
    \centering
    \subfigure{\includegraphics[width=8cm]{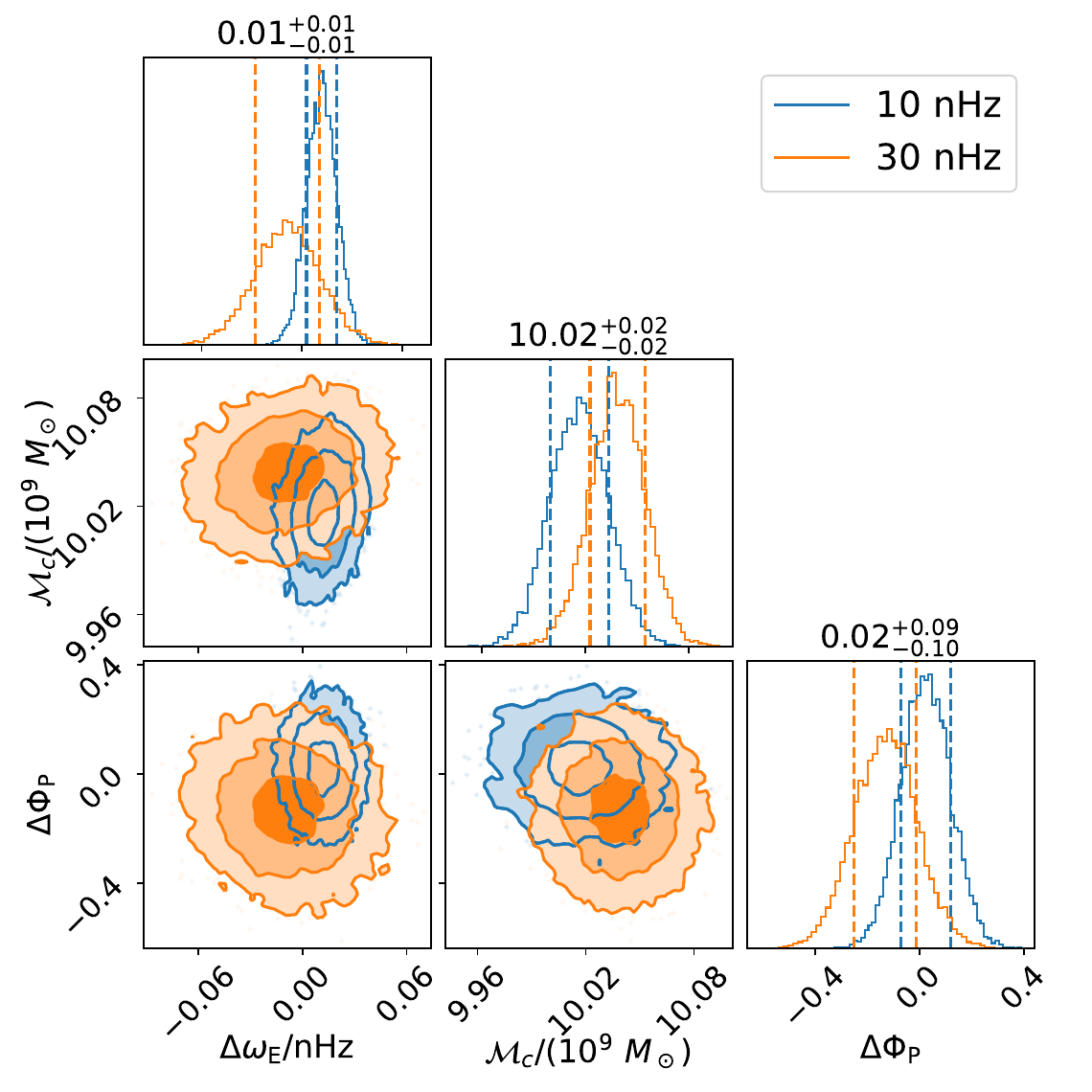}}
    \caption{The posterior corner plots of $\omega_\mathrm{E}$, $\mathcal{M}_c$ and $\Phi_\mathrm{P,J0613-0200}$. The blue (orange) contours represent the constraints
    inferred from PTA observations of a GW source with $\omega_\mathrm{E}=10 (30)$ nHz, $\mathcal{M}_c=10^{10}\ M_\odot$, $d_L=5$ Gpc, $T=30$ years. For convenient comparison, the differences with the injected values ($\Delta\omega_\mathrm{E}$ and $\Delta\Phi_\mathrm{P,J0613-0200}$) are plotted instead. Shown above each panel is the $1-\sigma$ error bar in the $\omega_\mathrm{E}=10$ nHz case.  }
    \label{fig:corner}
\end{figure}

In Fig. \ref{fig:corner}, we analyze two representative SMBHBs with a same chirp mass ($\mathcal{M}_c=10^{10}\ M_\odot$) and a same luminosity distance ($d_L=5$ Gpc), but different frequencies $\omega_{\rm E}=10 (30)$ nHz as examples. Assuming a mock PTA observation of $T=30$ years, we find their SNRs are $\mathrm{SNR}= 408$ and 263, respectively, which is consistent with the dependence of the signal amplitude on the frequency $R(t)\propto \omega_\mathrm{E}^{-1/3}$ [see Eq.~(\ref{eq_Rt})]. The posterior corner plots of three model parameters $\omega_{\rm E}, \mathcal{M}_{\rm c}, \Phi_{\rm P}$ are plotted,
where $\Delta \omega_{\rm E}/\omega_{\rm E}\sim \Delta \mathcal{M}_c/\mathcal{M}_c\sim \mathcal{O}(\mathrm{SNR}^{-1})$, 
and $\Phi_\mathrm{P,J0613-0200}$  is relevant to the signal of pulsar J0613−0200 only, 
therefore  is less constrained .

\begin{figure*}
    \centering
    \subfigure[J0030+0451, $T=10$ years]{\includegraphics[width=5.5cm]{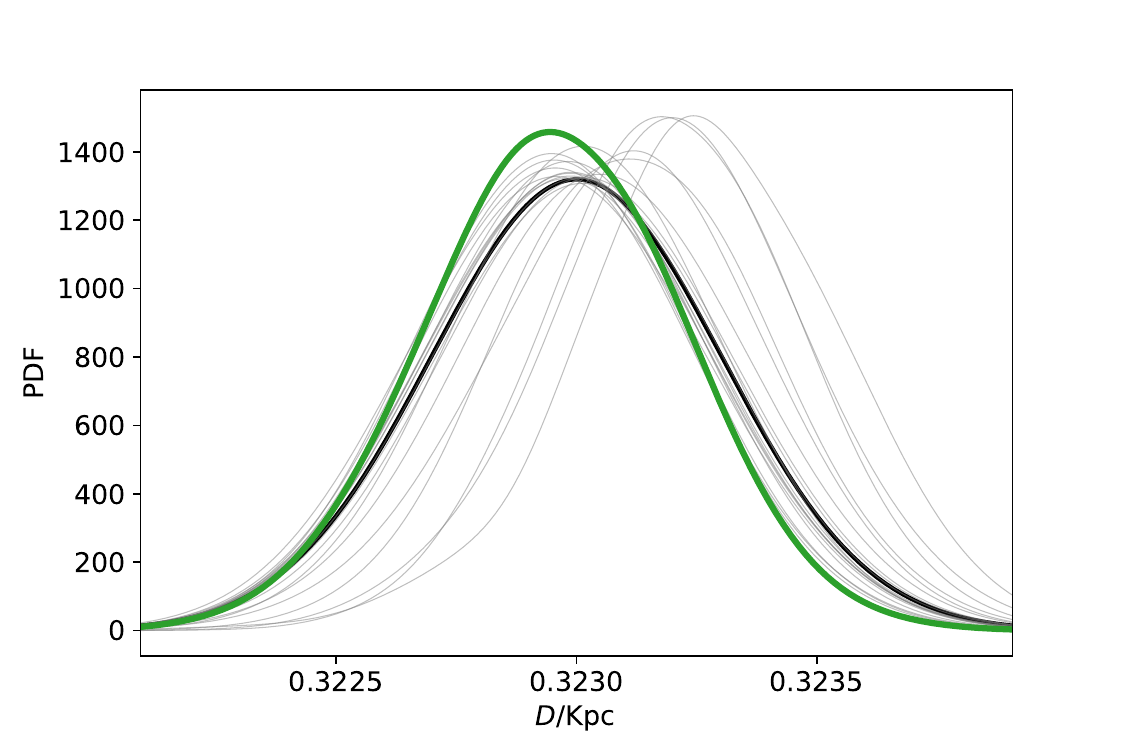}}
    \subfigure[J0613-0200, $T=10$ years]{\includegraphics[width=5.5cm]{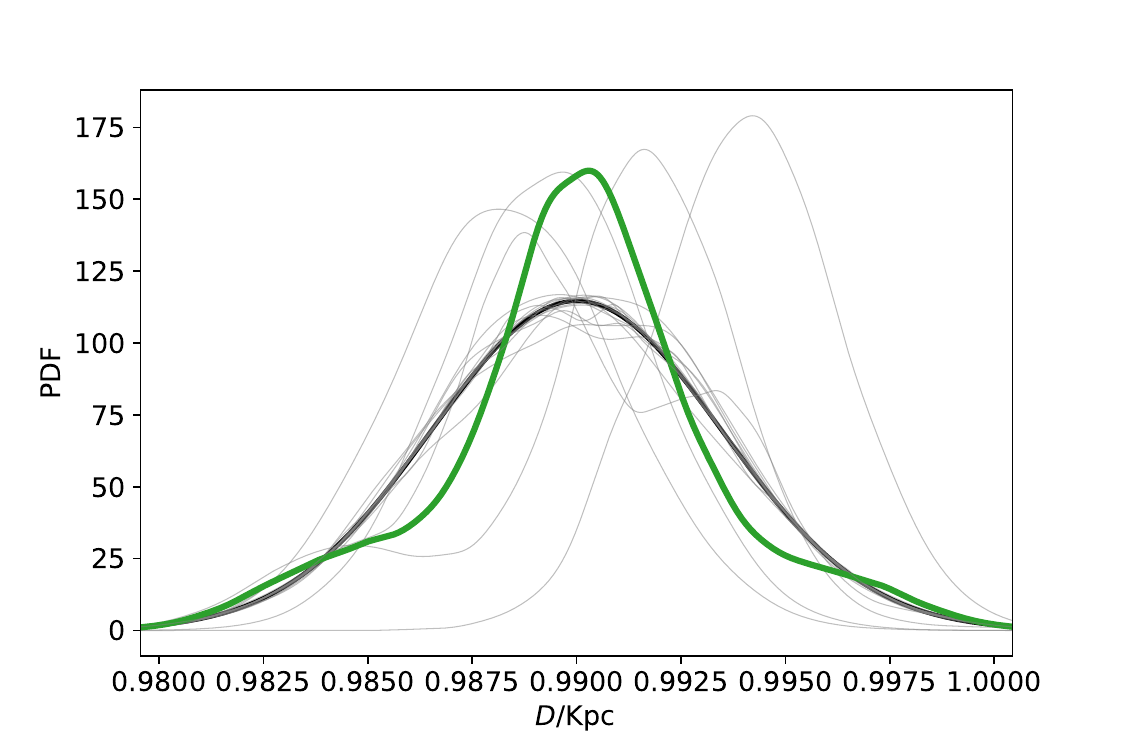}}
    \subfigure[J1911+1347, $T=10$ years]{\includegraphics[width=5.5cm]{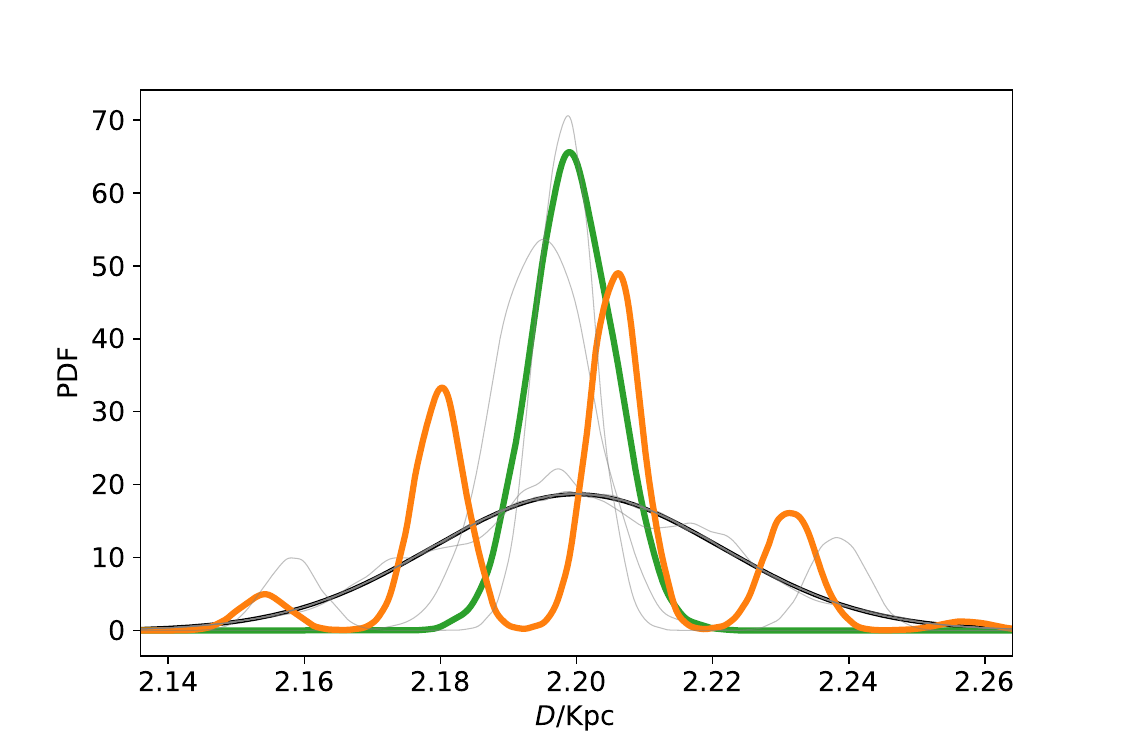}}
    \subfigure[J0030+0451, $T=30$ years]{\includegraphics[width=5.5cm]{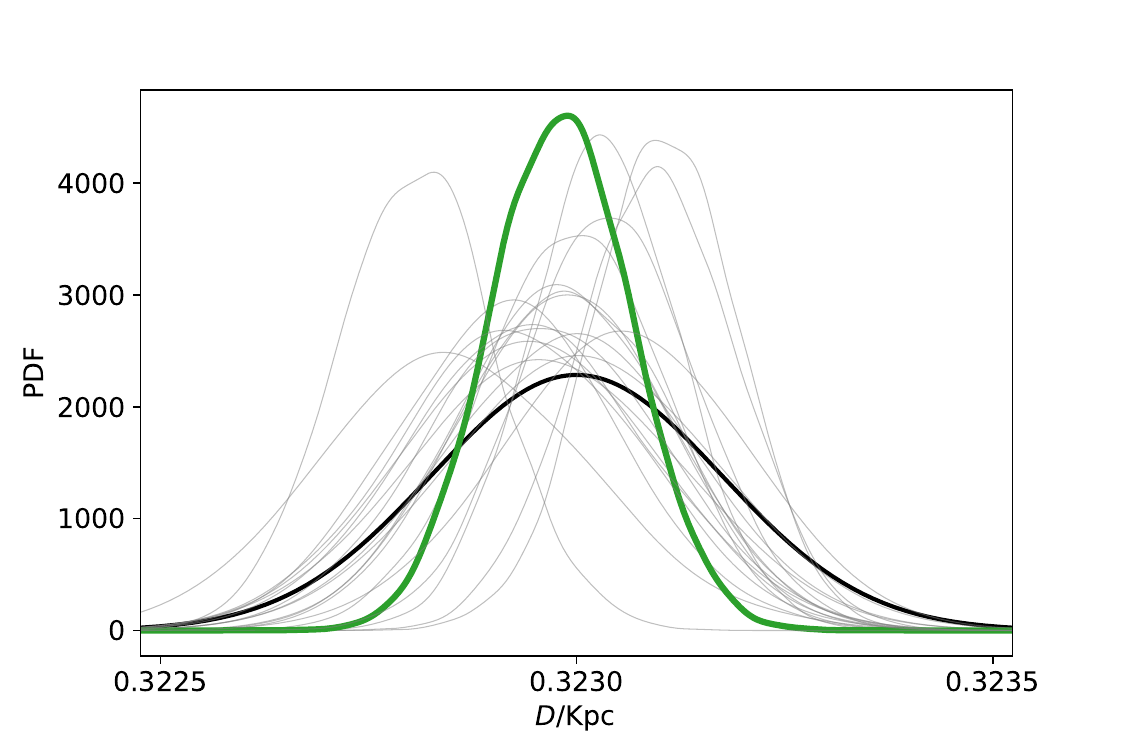}}
    \subfigure[J0613-0200, $T=30$ years]{\includegraphics[width=5.5cm]{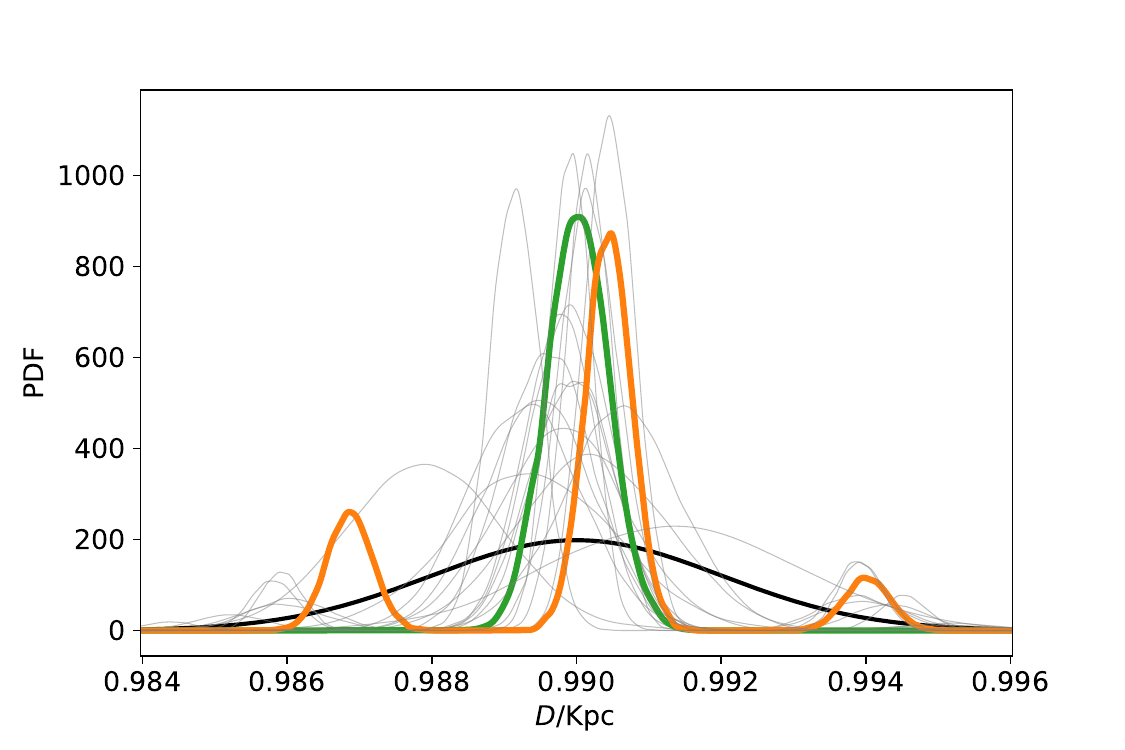}}
    \subfigure[J1911+1347, $T=30$ years]{\includegraphics[width=5.5cm]{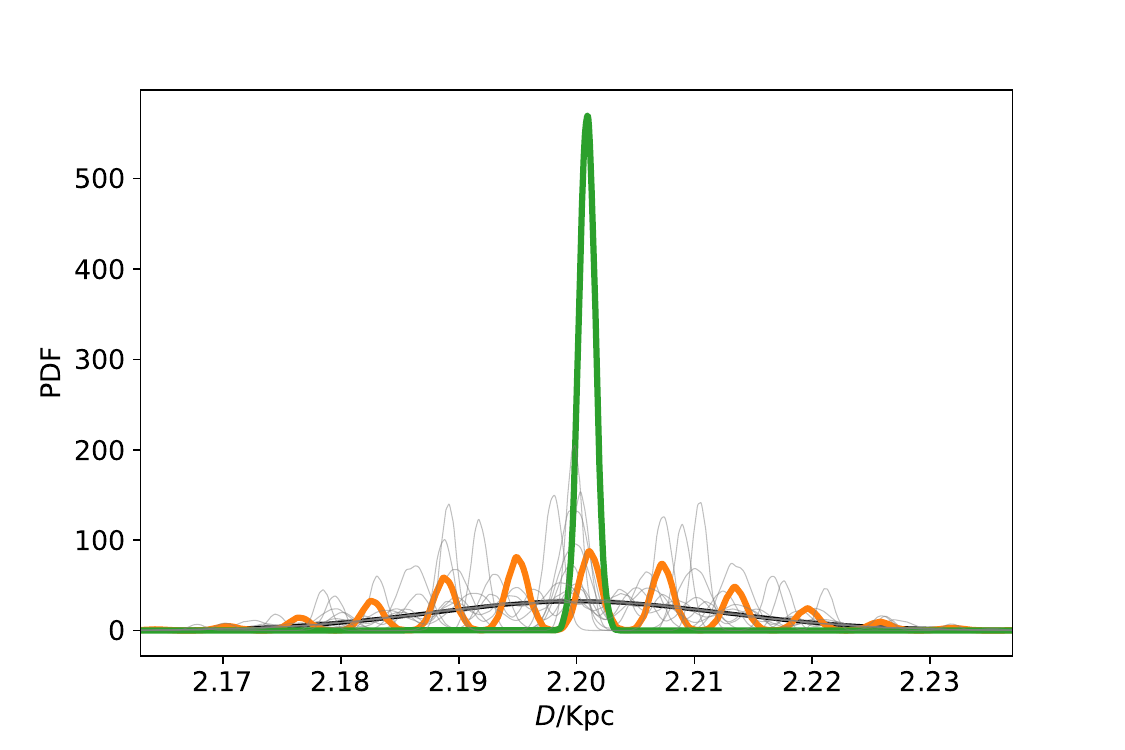}}
    \caption{The constraints of pulsar distance from SMBHBs with $\mathcal{M}_c=10^{10}\ M_\odot$, $\omega=10$ nHz, $d_L=5$ Gpc. The upper and bottom panels represent the results with $T=10,\ 30$ years' observation periods, respectively. And the left, middle, right panels represent the constraints on J0030+0451, J0613-0200, J1911+1347's distance, respectively. In each panel, a green line, an orange line and a black line are plotted, which represent a single peaked result, a multi-peaked result, and the timing parallax measurement, respectively. }
    \label{fig:D_constraints_10}
\end{figure*}

\begin{figure*}
    \centering
    \subfigure[J0030+0451, $T=10$ years]{\includegraphics[width=5.5cm]{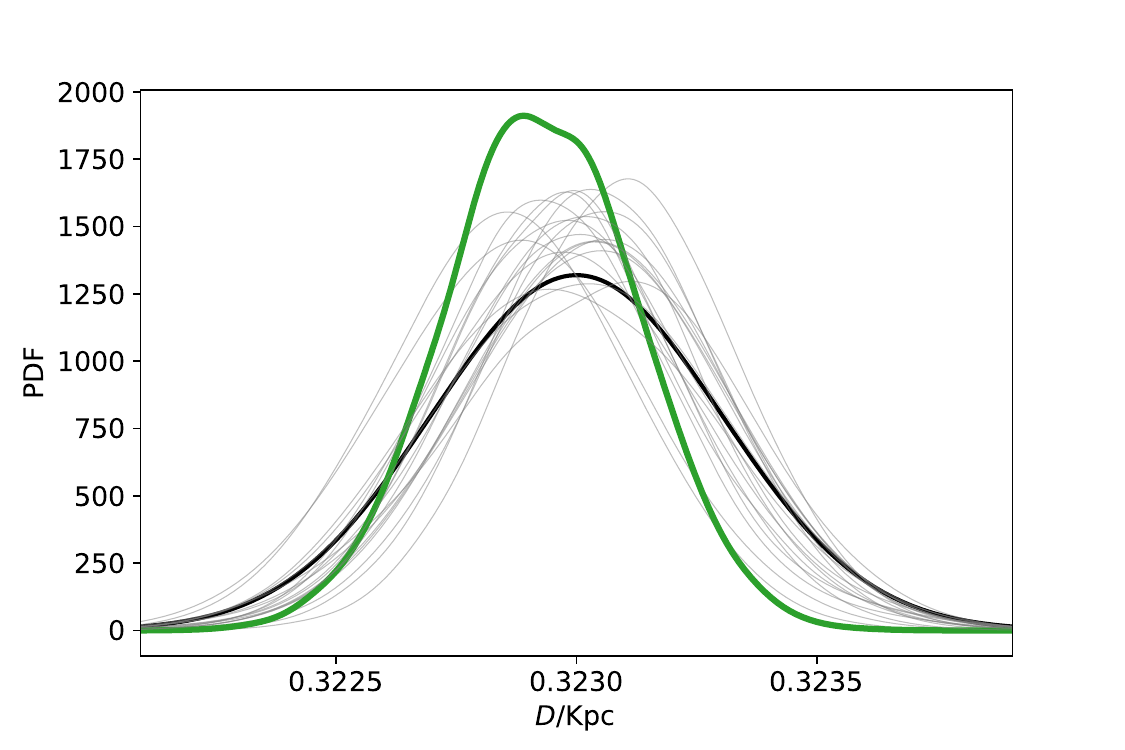}}
    \subfigure[J0613-0200, $T=10$ years]{\includegraphics[width=5.5cm]{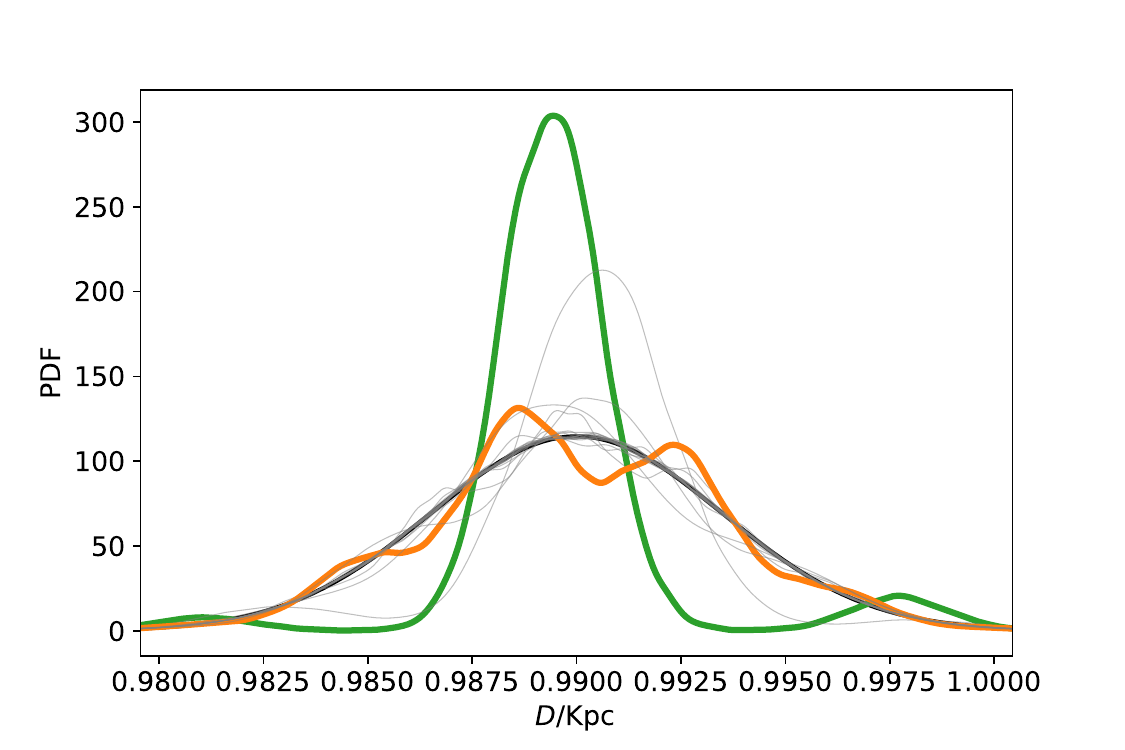}}
    \subfigure[J1911+1347, $T=10$ years]{\includegraphics[width=5.5cm]{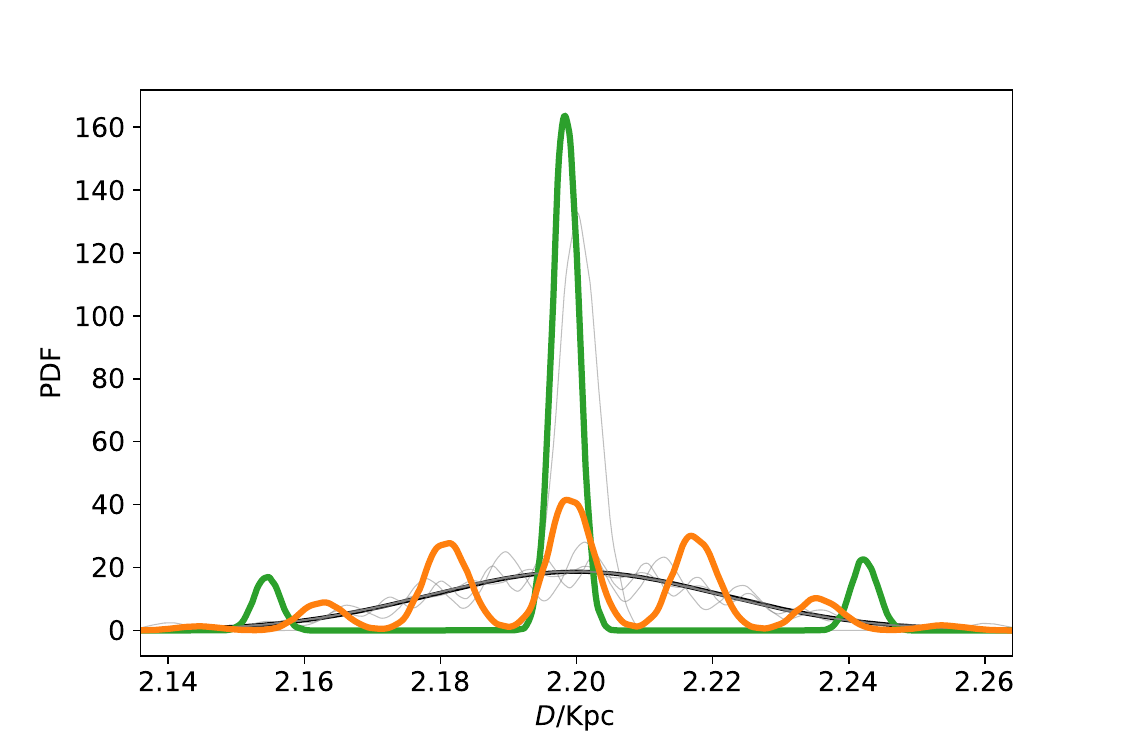}}
    \subfigure[J0030+0451, $T=30$ years]{\includegraphics[width=5.5cm]{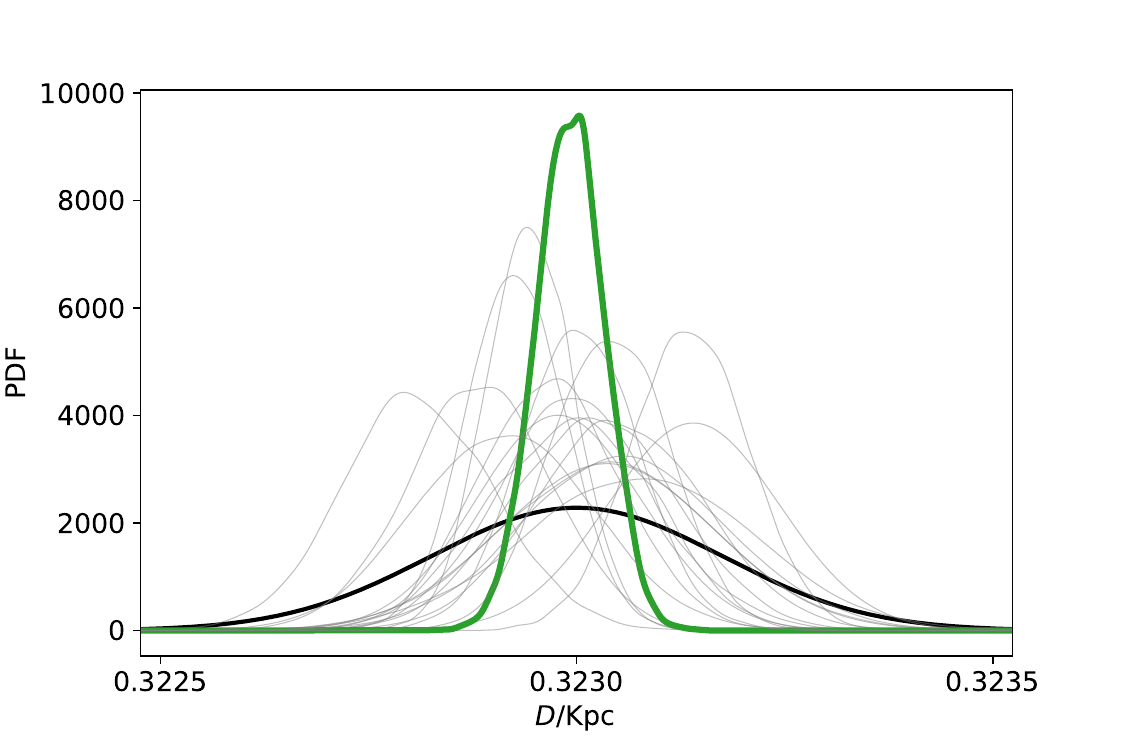}}
    \subfigure[J0613-0200, $T=30$ years]{\includegraphics[width=5.5cm]{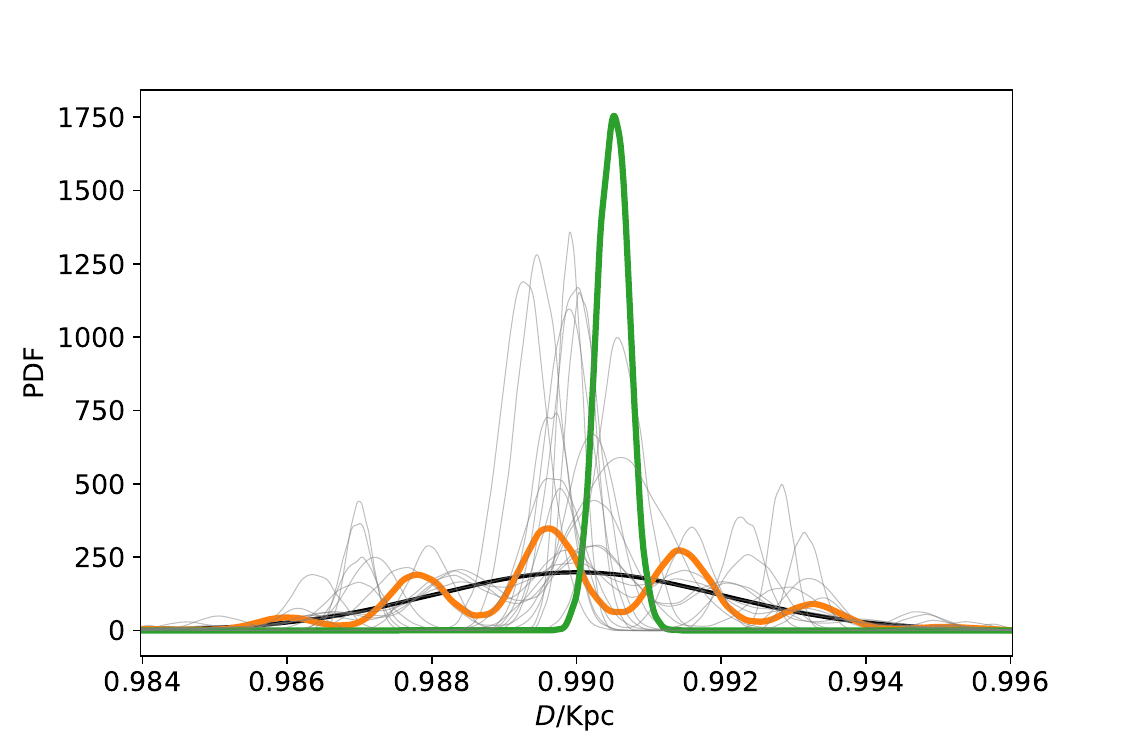}}
    \subfigure[J1911+1347, $T=30$ years]{\includegraphics[width=5.5cm]{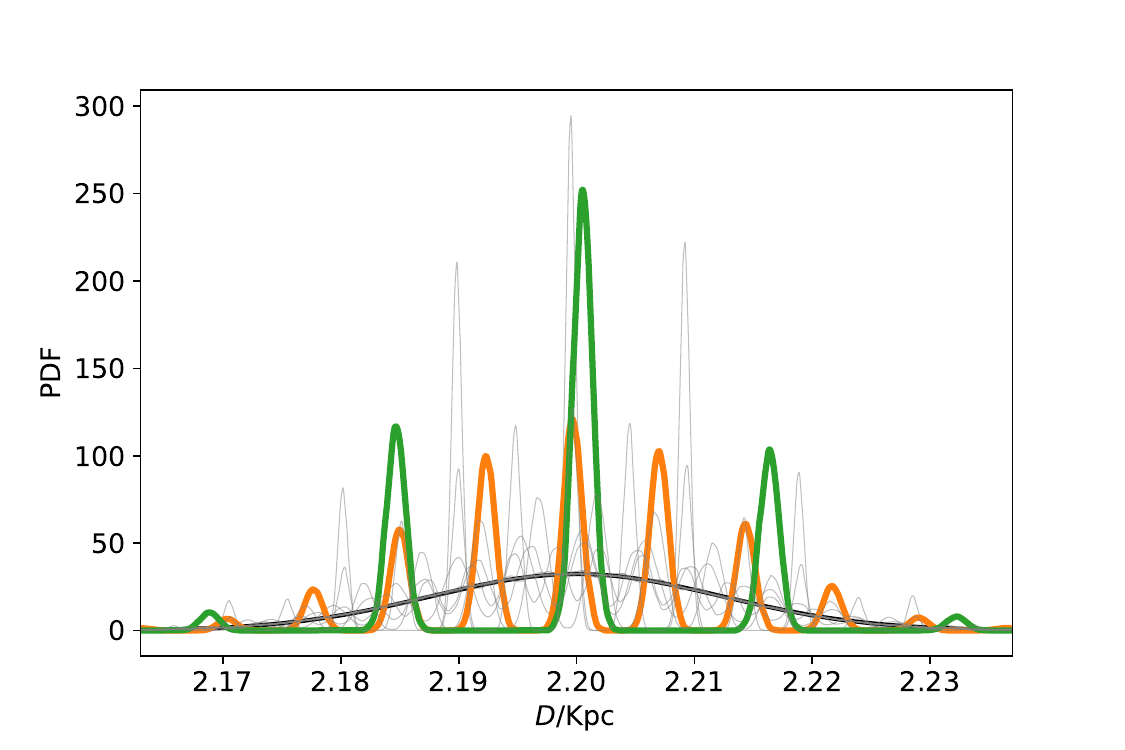}}
    \caption{The same with Fig. \ref{fig:D_constraints_10}, but for $\omega_\mathrm{E}=30$ nHz sources.}
    \label{fig:D_constraints_30}
\end{figure*}

\begin{figure*}
    \centering
    \subfigure[J0030+0451, $\omega_\mathrm{E}=10$ nHz]{\includegraphics[width=5.5cm]{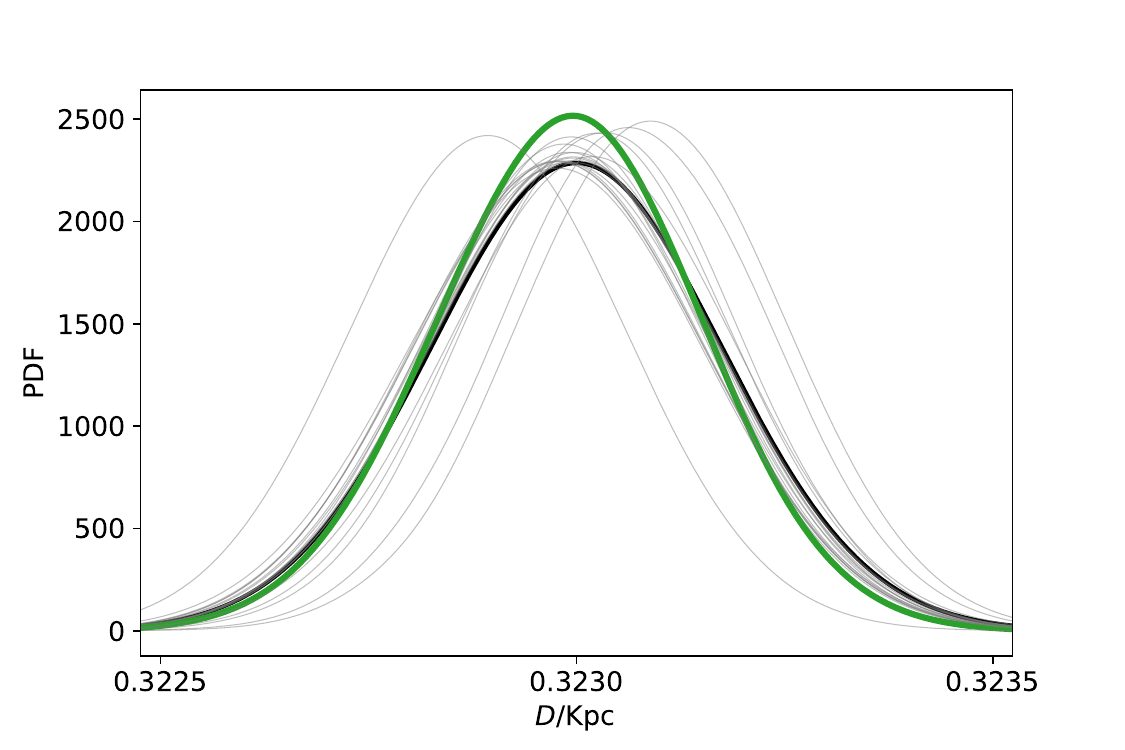}}
    \subfigure[J0613-0200, $\omega_\mathrm{E}=10$ nHz]{\includegraphics[width=5.5cm]{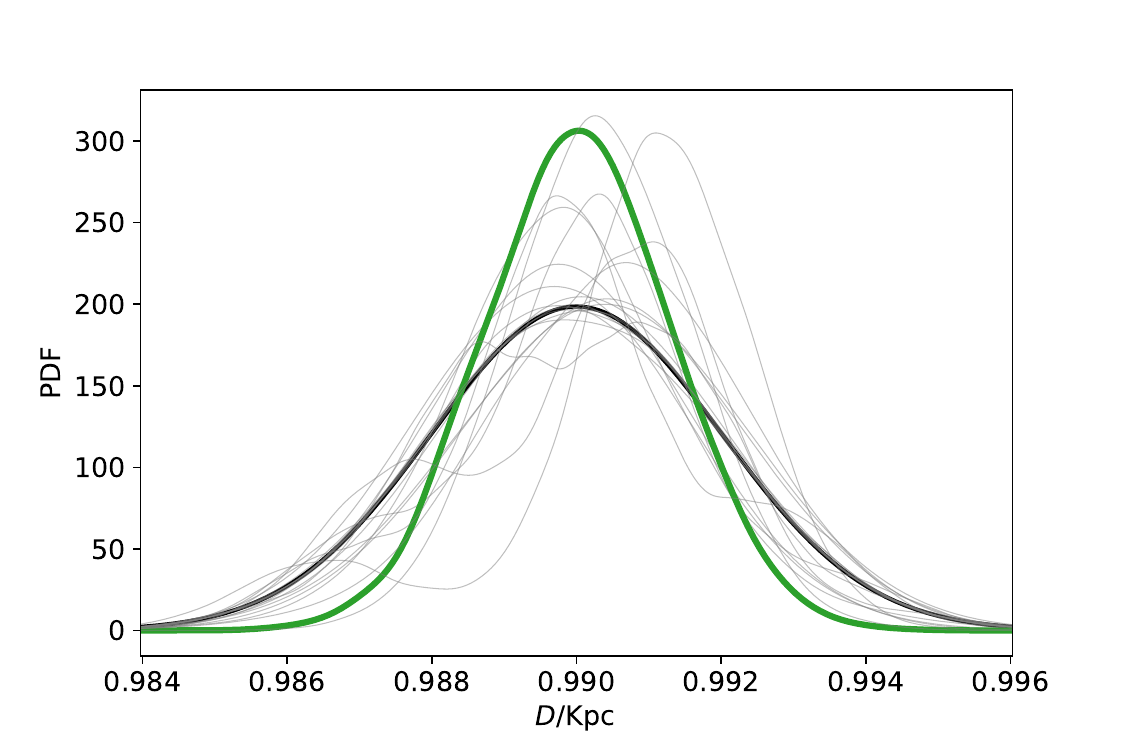}}
    \subfigure[J1911+1347, $\omega_\mathrm{E}=10$ nHz]{\includegraphics[width=5.5cm]{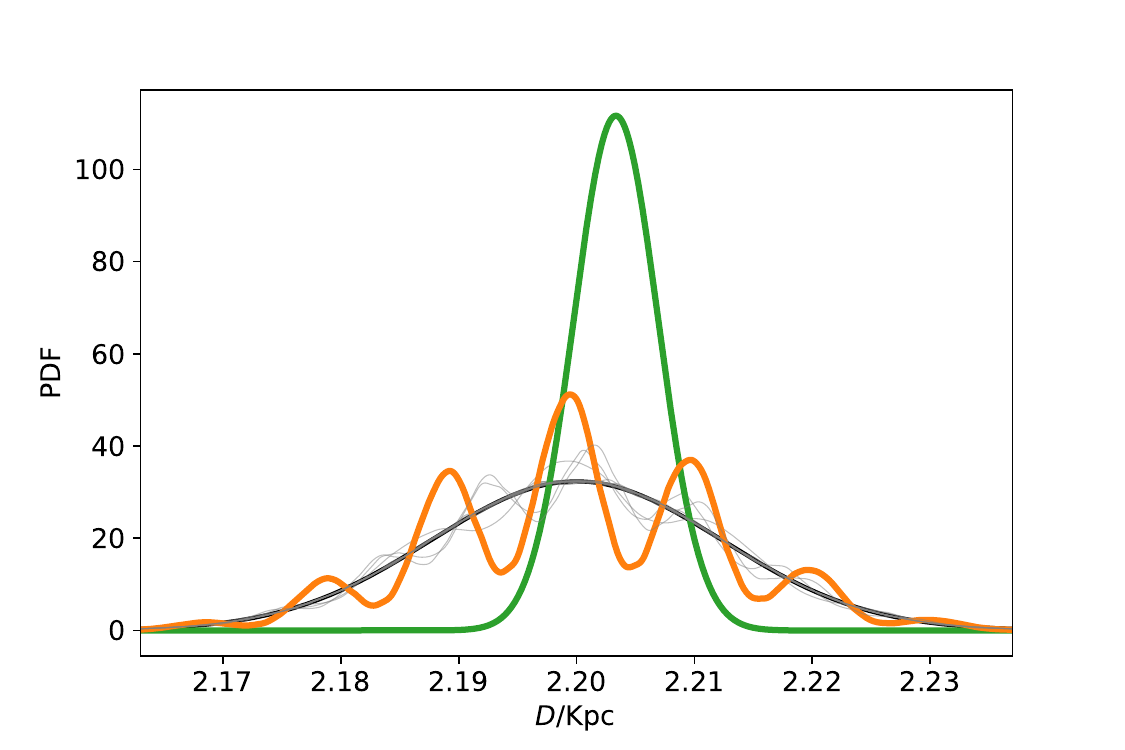}}
    \subfigure[J0030+0451, $\omega_\mathrm{E}=30$ nHz]{\includegraphics[width=5.5cm]{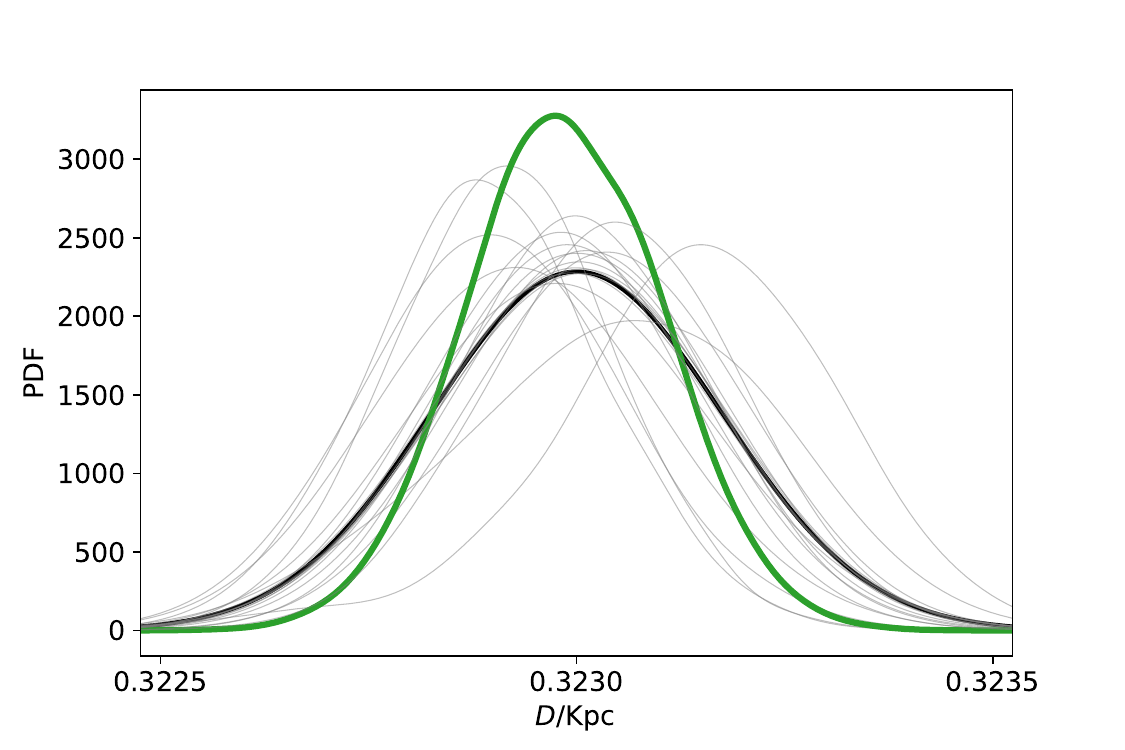}}
    \subfigure[J0613-0200, $\omega_\mathrm{E}=30$ nHz]{\includegraphics[width=5.5cm]{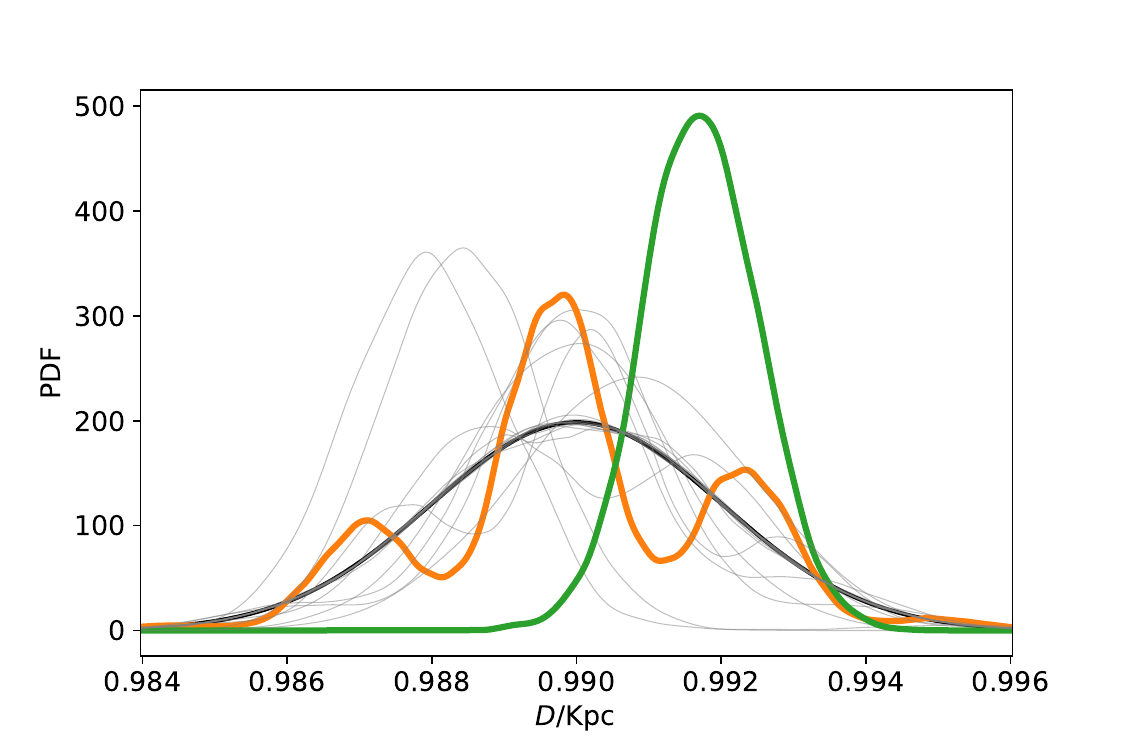}}
    \subfigure[J1911+1347, $\omega_\mathrm{E}=30$ nHz]{\includegraphics[width=5.5cm]{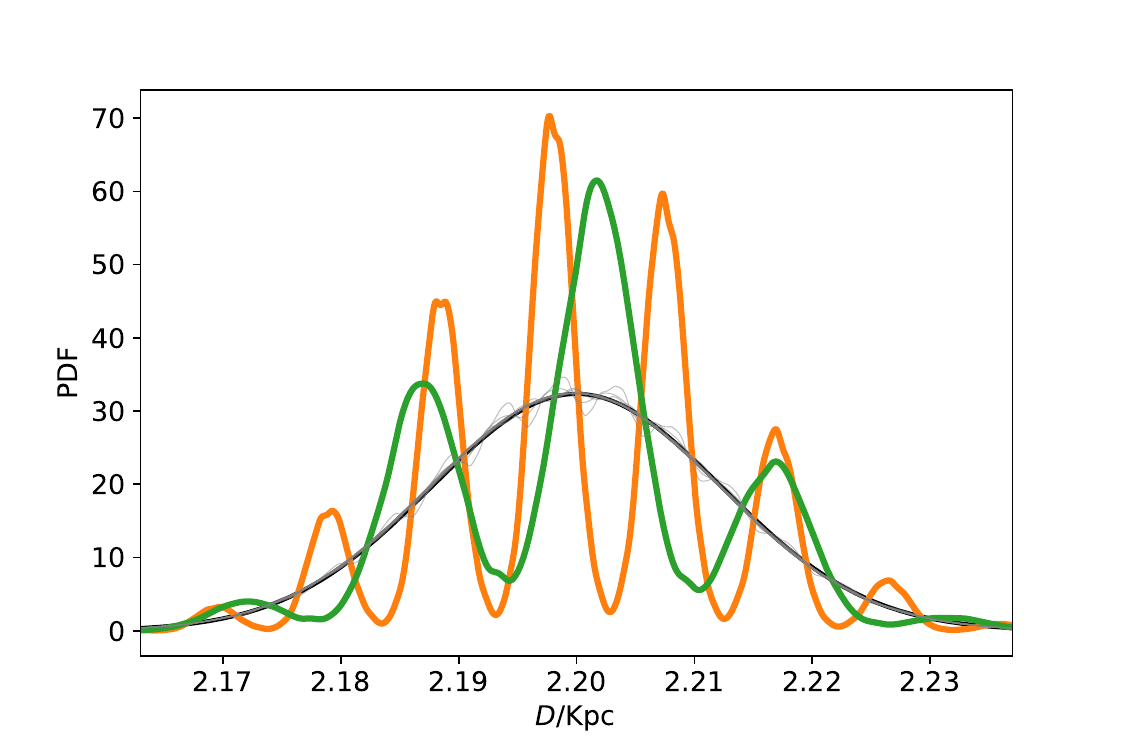}}
    \caption{Same to Fig. \ref{fig:D_constraints_10}, but for $d_L=20$ Gpc sources with $T=30$ years. The upper, bottom panels represent the constraints from $\omega_\mathrm{E}=10$ nHz, 30 nHz sources, respectively.  }
    \label{fig:D_constraints_dL20000}
\end{figure*}

Combining the prior information $\mathcal{\pi}(D)$ and the $\Phi_{\rm P}$ constraint, we obtain the pulsar distance constraint $\mathrm{PDF}(D_i)$ [Eq.~(\ref{eq:PD1})].
Fig. \ref{fig:D_constraints_10} and \ref{fig:D_constraints_30} show the constraints of pulsar distance produced by single events with $d_L=5$ Gpc. 
In each panel, the pulsar distance constraints from mock PTA observations of 20 SMBHBs are plotted with gray lines, as well as the timing parallax measurement plotted with a black line. To present the results more visually, we highlight a single peaked result with green line and a multi-peaked result with orange line.

From  Eqs.~(\ref{eq_sigma_D}) and (\ref{eq:golden}), a closer pulsar will have a higher fraction of golden events. Thus in the left panels, for the $D=0.323$ kpc pulsar, J0030+0451, all constraints are single peaked. In the case of $T=10$ yr, PTA observations of a SMBHB with $\omega_\mathrm{E}=10$ nHz can slightly improve the constraint on $D$. The best constraint is $D=323.18^{+0.24}_{-0.24}$ pc, which is 13\% tighter than the timing parallax measurement. For $\omega_\mathrm{E}=30$ nHz events, the constraints are generally better than those at lower frequency. As a result, the best measurement of $D$ reaches about $\pm0.20$ pc. 

With longer observation time span $T$, the uncertainty of timing parallax measurement scales as $T^{-1/2}$, but the uncertainties of $\omega_\mathrm{E}$ and $\Phi_\mathrm{P}$ are roughly proportional to $T^{-1}$. So in the bottom panels, the improvements produced by the PTA observations are much larger than in the $T=10$ years case. In the case with $\omega_\mathrm{E}=30$ nHz, $T=30$ years, the best constraint reaches $0.04$ pc, which improve the timing parallax measurement uncertainty 
$\sigma_D=0.17$ pc  by a factor of 4. 

When the pulsar is farther away, the periodicity in the pulsar distance posterior becomes harder to eliminate due to the worse timing parallax prior. Thus, in the case of $\omega_\mathrm{E}=30$ nHz with $D=0.99$ kpc pulsar J0613-0200, 
there are multiple peaks in the posterior. As for those golden events, the best constraint reaches $\sim0.2$ pc. For the $D=2.2$ kpc pulsar J1911+1347, 
only nearly aligned SMBHBs with  $\cos\theta\sim1$ satisfy the criterion for golden events. Therefore, the fractions of golden events are much smaller than for closer pulsars, as shown in the right panels of the Fig. \ref{fig:D_constraints_10} and \ref{fig:D_constraints_30}. Despite their rarity, these golden events can all significantly improve the constraints from $\sim10$ pc to a few or sub pc.

 For more distant SMBHBs with $d_{\rm L}=20$ Gpc, the SNRs are much lower and we only show the pulsar distance measurement results with long observation span  $T=30$ years in Fig.~\ref{fig:D_constraints_dL20000}. From the left panels, for J0030+0451, a $\omega_\mathrm{E}=10$ nHz event cannot effectively improve the measurement of $D$. However, several high frequency SMBHBs can still constrain $D$ to $\lesssim0.14$ pc. In the upper middle panel, for J0613-0200, 8 (7) events 
qualify as golden events in the case of $10 (30)$ nHz, respectively. The best constraints for  for J0613-0200 are $D=990.2^{+1.3}_{-1.3}$ pc and $D=991.6^{+0.8}_{-0.8}$ pc in the two cases, respectively. 
For J1911+1347, only one SMBHB with  $\omega_\mathrm{E}=10$ nHz qualifies a golden event, from which the pulsar distance   is constrained as $D=2203.1^{+3.6}_{-3.5}$ pc.

From the results above, we can conclude that for a $D\lesssim1$ kpc pulsar, nHz GW observations of SMBHBs in combination with the timing parallax measurement are highly likely to constrain the pulsar distance to a sub-pc level. 
For more distance pulsars, golden events are possible only when the source's direction is is closely aligned with the pulsar.  However, considering the limited precision of parallax measurements for these pulsars,  a single golden event could significantly improve the pulsar distance measurements.

\subsection{Constraints from Multiple Sources}
In the previous subsection, we show the constraint produced by each  golden event alone. In addition to those golden events, there are a number of events that cannot individually constrain pulsar distance due to the high periodicity in the posteriors. 
It is interesting to note that the peaks in the posteriors tend to be rather narrow (see the orange lines in Fig. \ref{fig:D_constraints_30} as examples). As mentioned in Sec. \ref{sec:methodology}, if the `fake' peaks can be eliminated by other 
GW sources with different periods, these events can provide a good joint constraint on the pulsar distance.

\begin{figure}
    \centering
    \subfigure[J1911+1347, $\omega_\mathrm{E}=30$ nHz]{\includegraphics[width=8cm]{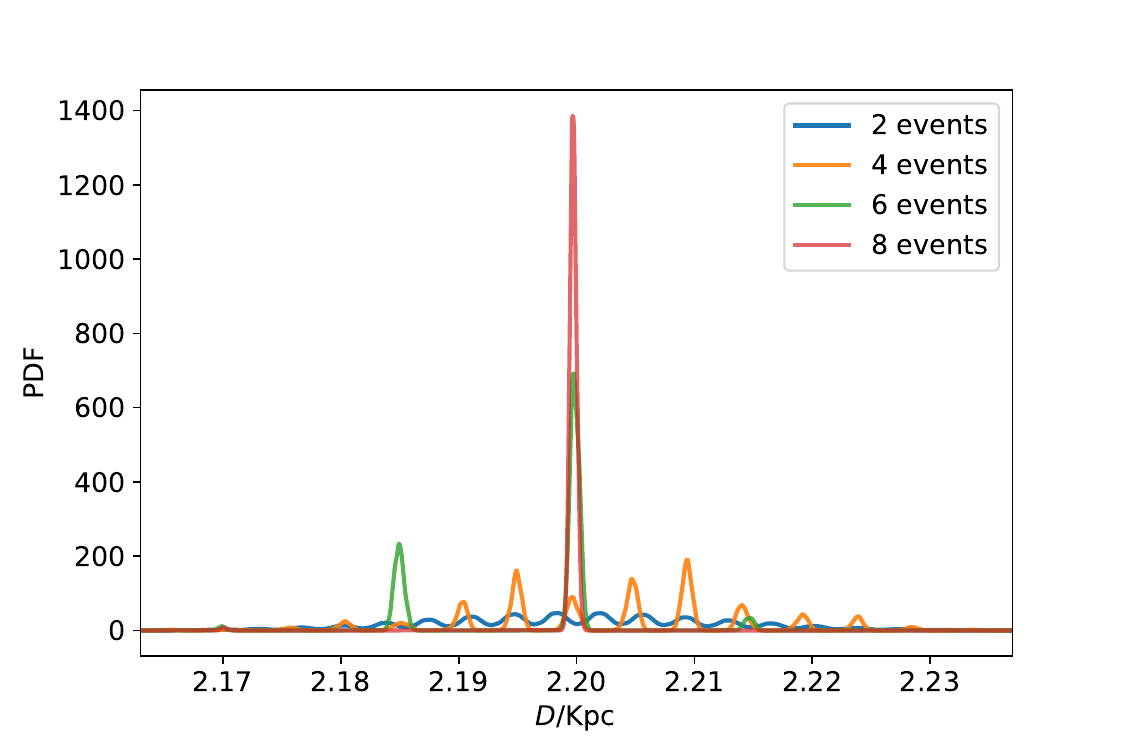}}
    \caption{We select the multi-peaked posteriors of J1911+1347's distance given by $\omega_\mathrm{E}=30$ nHz, $T=30$ years, $d_L=5$ Gpc sources, and show the joint constraints of these posteriors with different numbers of events.}
    \label{fig:multi}
\end{figure}

To show the feasibility of this method, we select the J1911+1347's distance posteriors with multi-peaked structure from the lower-right panel of Fig. \ref{fig:D_constraints_30}, and show their joint constraints with different event numbers in Fig. \ref{fig:multi}. When using $\leq$4 SMBHBs, the combined results still show strong periodicity. As more SMBHBs are included, false peaks in the posteriors gradually disappear. With 8 SMBHBs, only one clear peak remains, giving a pulsar distance constraint of $\pm$0.28 pc. 

\begin{figure*}
    \centering
    \subfigure[J0030+0451, $d_L=5$ Gpc]{\includegraphics[width=5.5cm]{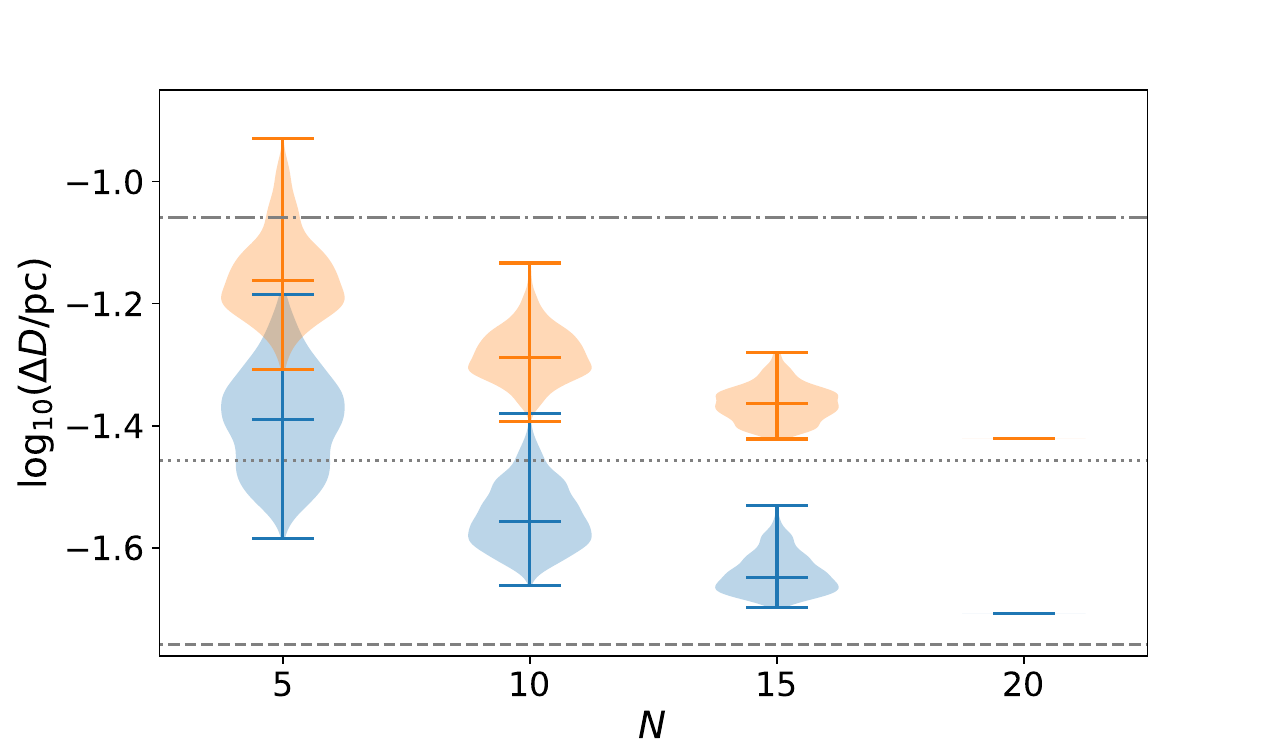}}
    \subfigure[J0613-0200, $d_L=5$ Gpc]{\includegraphics[width=5.5cm]{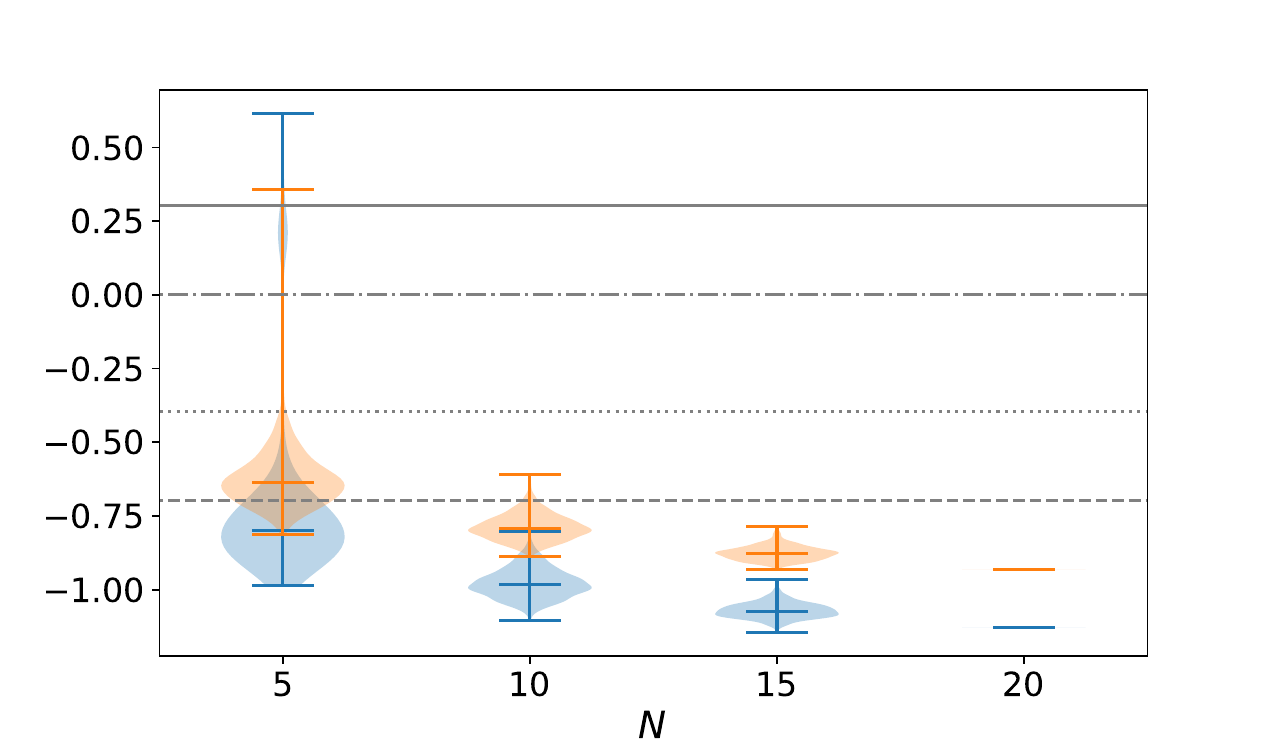}}
    \subfigure[J1911+1347, $d_L=5$ Gpc]{\includegraphics[width=5.5cm]{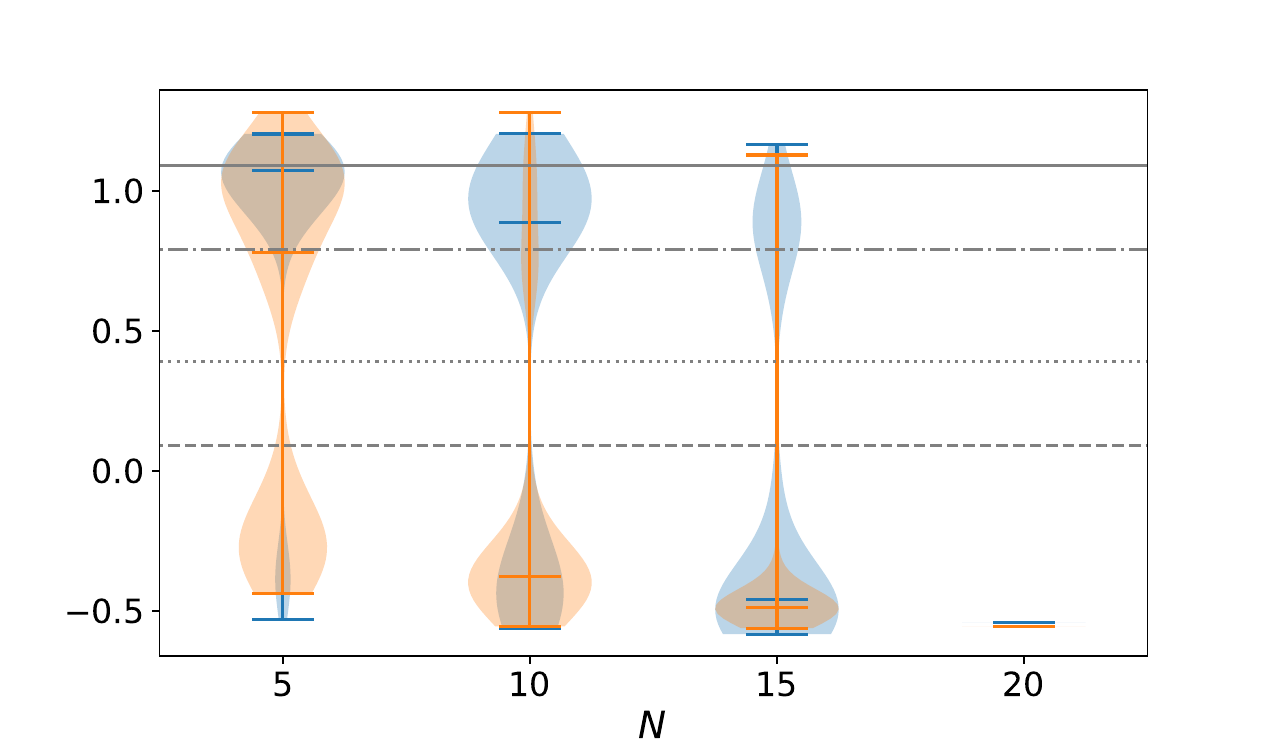}}
    \subfigure[J0030+0451, $d_L=20$ Gpc]{\includegraphics[width=5.5cm]{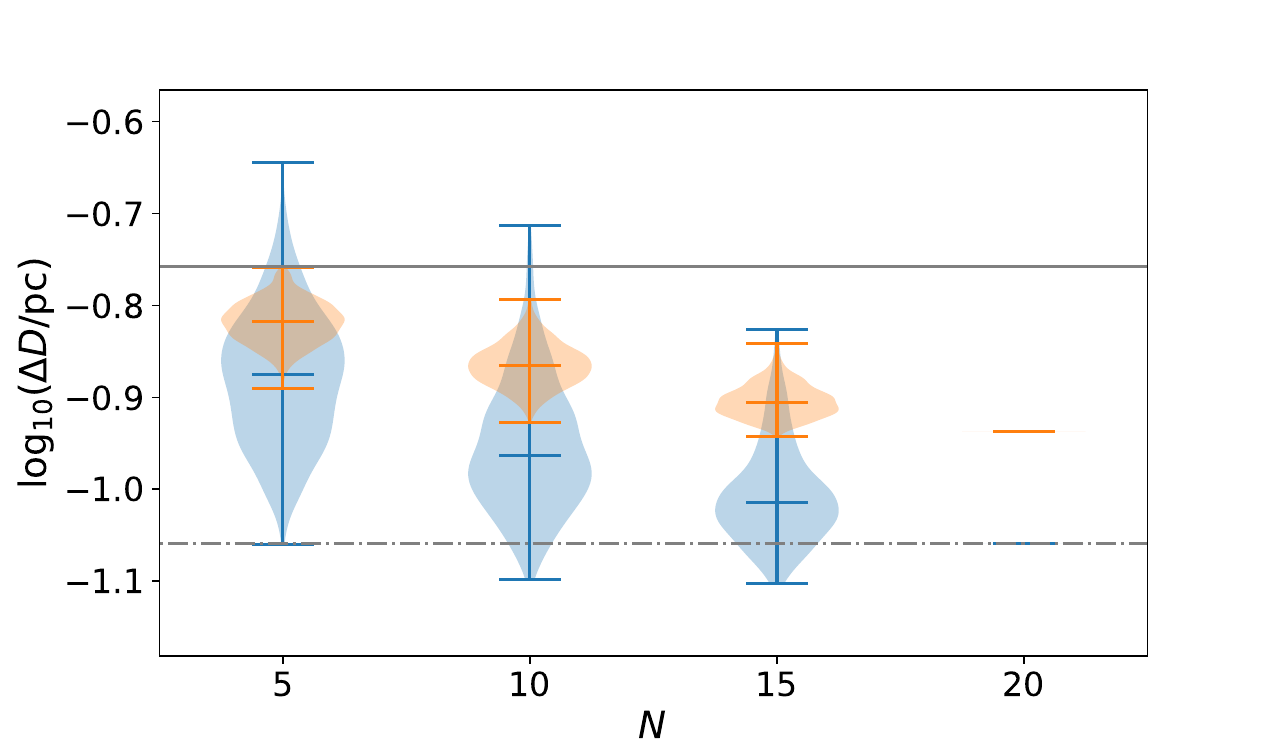}}
    \subfigure[J0613-0200, $d_L=20$ Gpc]{\includegraphics[width=5.5cm]{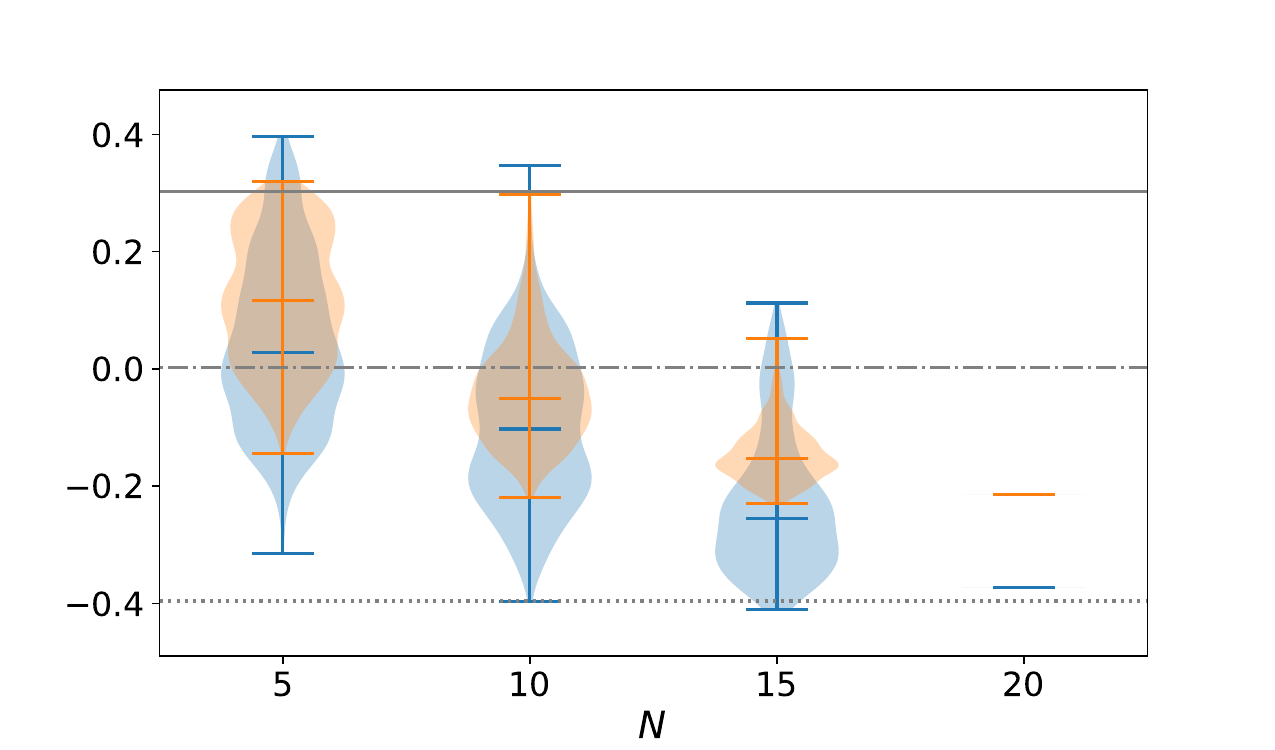}}
    \subfigure[J1911+1347, $d_L=20$ Gpc]{\includegraphics[width=5.5cm]{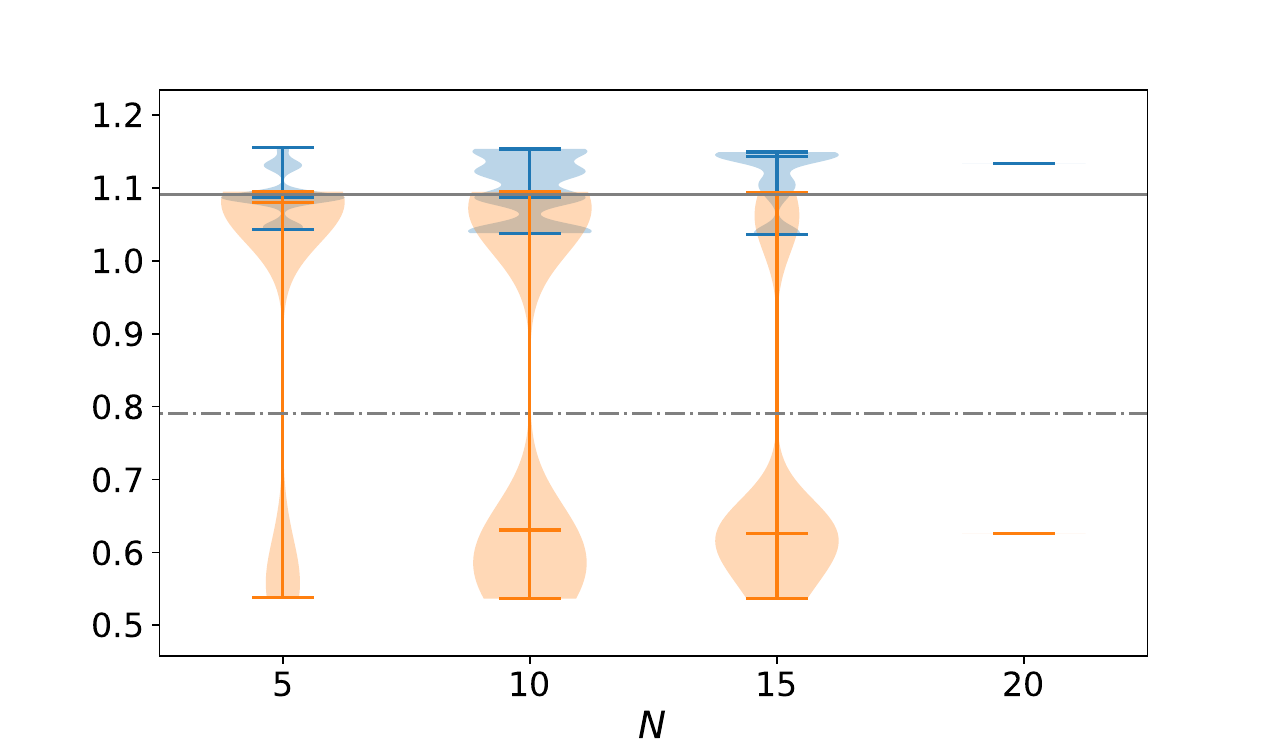}}
    \caption{Distributions of pulsar distance's constraints $\Delta D$ with different numbers  $N$ of GW sources. The orange and blue violins represent the distributions of $\omega=10$ nHz and $\omega=30$ nHz sources, respectively. The observation time spans for all sources are 30 years. As a reference, the timing parallax measurements $\sigma_D$ as well as $0.5\sigma_D$, $0.2\sigma_D$, $0.1\sigma_D$ are also plotted in solid, dashdot, dotted, dashed gray lines, respectively. 
    }
    \label{fig:D_constraints_violin}
\end{figure*}

In the example above, the true peak stands out with fake peaks eliminated from multiple non-golden events. In actual observations, the inclusion of golden events will significantly accelerate this process.
As a result, the joint constraint from multiple sources  is highly sensitive to the existence of golden events. To demonstrate the joint constraints from different SMBHB samples, we perform 1000 repeated random samplings of 5, 10, and 15 subsets from each of the 20 previously simulated SMBHB populations. Fig. \ref{fig:D_constraints_violin} shows the derived constraint distributions on pulsar distances for these subsamples under a 30-year observation period. As a reference, the timing parallax measurement $\sigma_D$ as well as $0.5\sigma_D$, $0.2\sigma_D$, $0.1\sigma_D$ are plotted in each panel. 

For J0030+0451, we find the pulsar distance constraint from multiple sources scale as  $\Delta D\propto N^{-1/2}$ because all events are golden events. Generally, the results of $\omega_\mathrm{E}=30$ nHz sources are better than the $\omega_\mathrm{E}=10$ nHz cases due to better measurement of $\Phi_\mathrm{P}$. For both cases with $d_L=5$ Gpc, $>90\%$ sub-samples of $N=5$ can constrain  the pulsar distance with uncertainty $<0.5\sigma_D$. And 10 sources with $\omega_\mathrm{E}=30$ nHz, $d_L=5$ Gpc can constrain $D$ to about $\pm0.03$ pc. In the case of $d_L=20$ Gpc, only a small fraction of $N=15$, $\omega_\mathrm{E}=30$ nHz sub-samples reach an accuracy of $0.5\sigma_D$.

For J0613-0200, as the fraction of golden events is still high, the constraints are also roughly $\propto N^{-1/2}$. Only a few sub-samples with $N=5$ still have a multi-peaked measurement of $D$, and their results are much worse than other sub-samples and approach $\sigma_D$. Since the J0613-0200's $\sigma_D$ is much larger than J0030+0451's, 10 $d_L=5$ Gpc sources can improve the pulsar distance measurements to $0.1\sigma_D$ in both high frequency and low frequency cases. And when $d_L=20$ Gpc, the constraints can also reach $0.5\sigma_D$.

For the J1911+1347, the high periodicity leads to a clear stratification in the pulsar distance constraints. The upper part consists of sub-samples where false peaks are not completely eliminated, and the constraints are typically worse the 4 pc. In contrast, the constraints are significantly better  with $\Delta D\lesssim1$ pc in the lower part. 

In the case of $d_L=20$ Gpc, $\omega_\mathrm{E}=10$ nHz,  the golden event (the green line in the upper right panel of Fig. \ref{fig:D_constraints_dL20000}) improves the distance constraint $D$ to $\sim4$ pc
though the joint constraint from all the non-golden events does not contribute much constraining power. 
In the case of $\omega_\mathrm{E}=30$ nHz, no golden events exist, and the combined constraints from all non-golden events still exhibit multiple peaks, failing to enhance distance measurement precision.

\subsection{Constraints with Multiple Sources: a realistic population}

To demonstrate the actual use of nHz GWs in measuring pulsar distances, we sample SMBHBs from the BBH population model constructed by \cite{2020ApJ...897...86C} and \cite{2023SCPMA..6620402B}. 
First, we adopt the mass distribution model in \cite{2023SCPMA..6620402B}, which is based on the NANOGrav 15-year data set (See Fig. 3 in \cite{2023SCPMA..6620402B}) for details). For SMBHBs with large chirp mass ($M_c>10^9$), the SMBHB merger rate $R$ is roughly power-law distributed with $M_c$ with an index of -3.5, where $M_c=\mathcal{M}_c/(1+z)$ is the physical chirp mass. For the redshift distribution, we adopt the model in \cite{2023SCPMA..6620402B}, which combines various BBH evolution tracks, galaxy mass functions, galaxy merger rates and SMBH-host relations. The Fig. 17 in \cite{2020ApJ...897...86C} shows that for massive SMBHBs, $\log_{10}R$ is roughly proportional to $z^{-0.4}$. After combining their models, the merge rate of SMBHB with $M_c\in[10^{9},10^{10.5})\ M_\odot$ and $z\in[0.01,3)$ is about $1.2\times10^{-3}$ year$^{-1}$. 

\begin{figure}
    \centering
    \includegraphics[width=8cm]{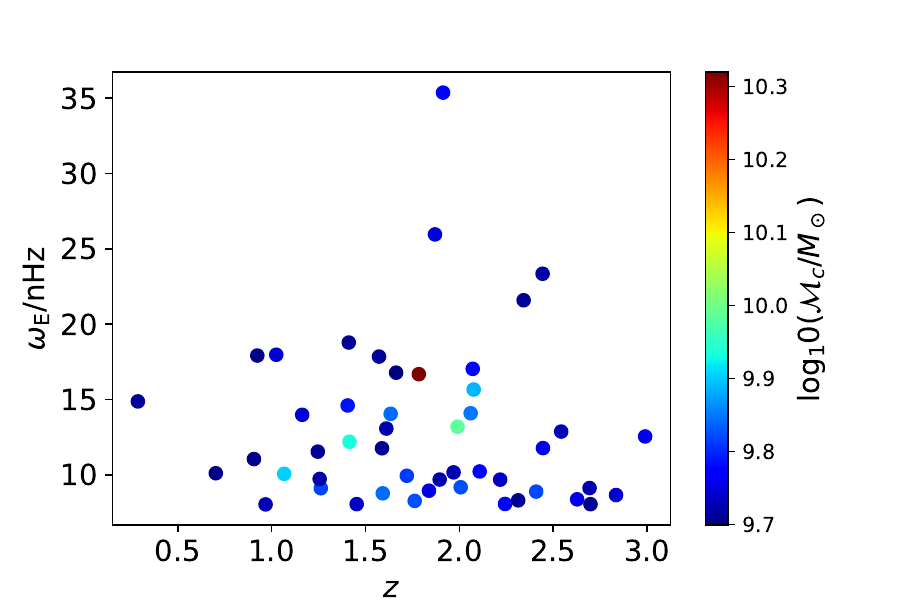}
    \caption{The parameters of 48 SMBHB samples simulated, where the $x$-axis, $y$-axis, color represent  source redshift $z$, source frequency $\omega_\mathrm{E}$ and chirp mass $\mathcal{M}_c$, respectively.}
    \label{fig:sample}
\end{figure}

From Eq.~(\ref{eq_domega}), it is straightforward to find the time to coalescence is 
\begin{equation}
    t-t_c=\frac{5}{2^{16/3}}\mathcal{M}_c^{-5/3}\omega^{-8/3}\, 
\end{equation}
where $t-t_c\sim$ Myr for  a nHz source with $\mathcal{M}_c=10^{10}\ M_\odot$. 
The number of sources with GW frequencies lying between $\omega_1$ and $\omega_2$ is therefore $R\times[t(\omega_1)-t(\omega_2)]$. 
As a result, we find $10^4$ sources in total with $\omega\in[5,50)$ nHz, $M_c\in[10^{9},10^{10.5})\ M_\odot$ and $z\in[0.01,3)$. 
We randomly sample $10^4$ SMBHBs from the distribution above and select massive sources with $\mathcal{M}_c>5\times10^9\ M_\odot,\ \omega_\mathrm{E}>8$ nHz (approximately the lowest frequency a PTA can probe within 30 years). 
48 SMBHBs satisfying the above criteria are selected and are shown in Fig. \ref{fig:sample}. 

\begin{figure*}
    \centering
    \subfigure[J0030+0451]{\includegraphics[width=8cm]{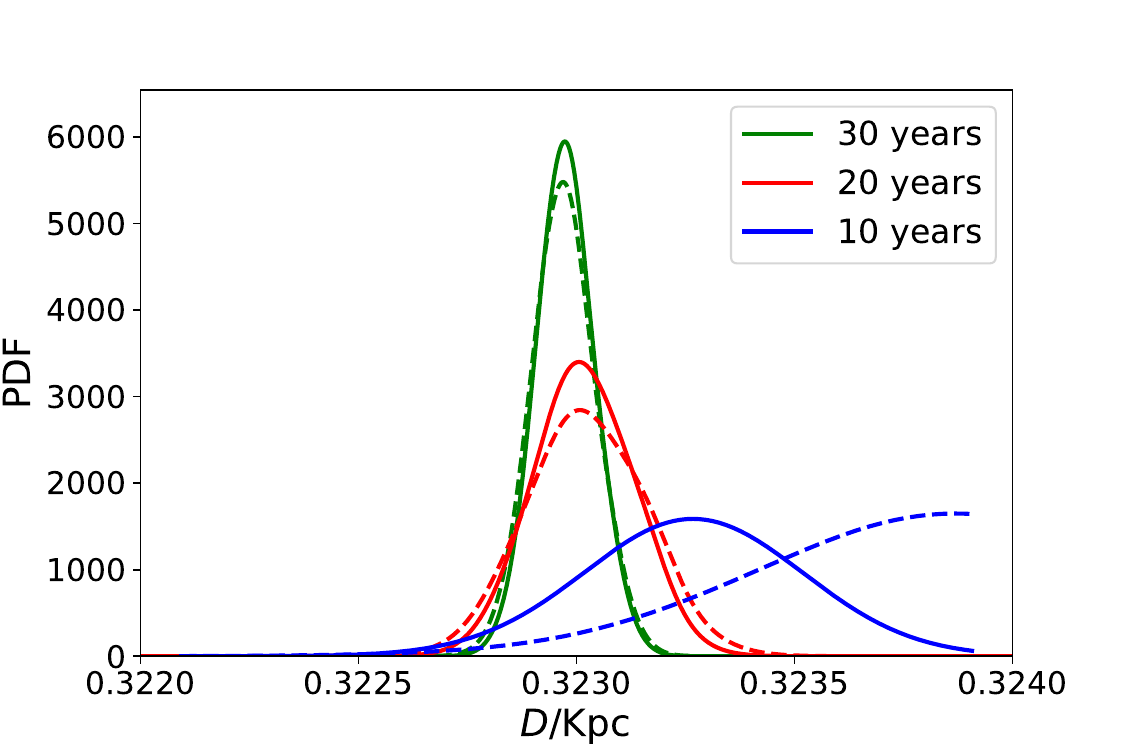}}
    \subfigure[J0613-0200]{\includegraphics[width=8cm]{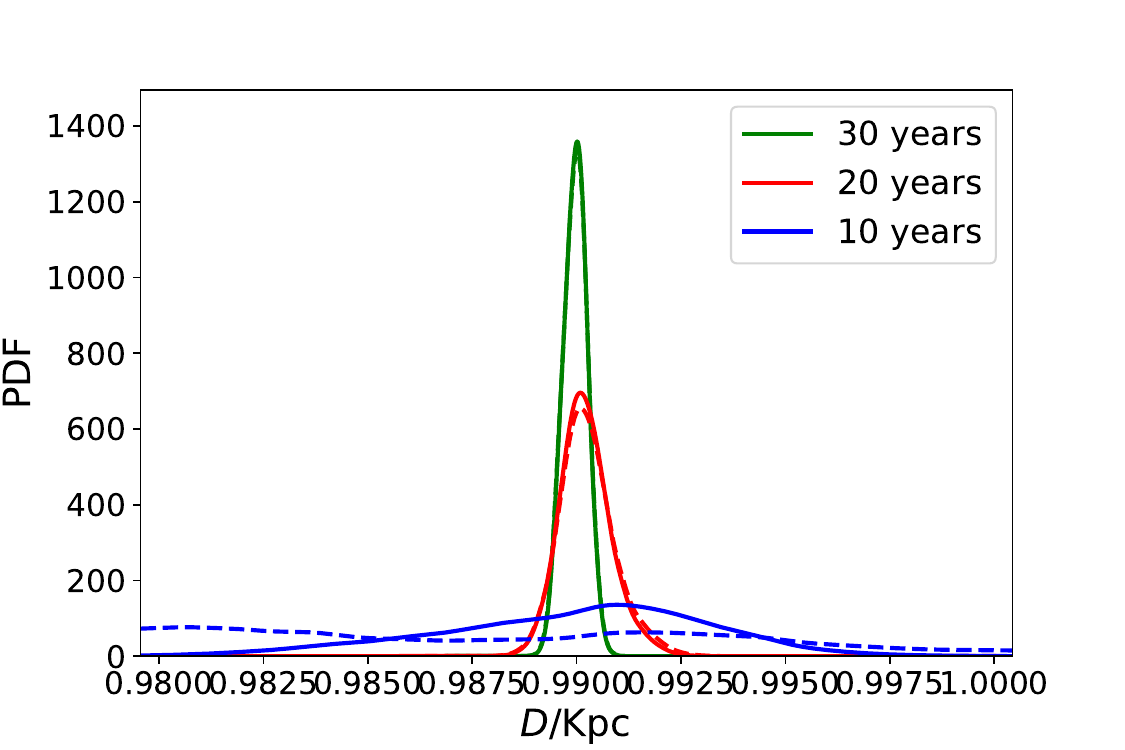}}
    \subfigure[J0751+1807]{\includegraphics[width=8cm]{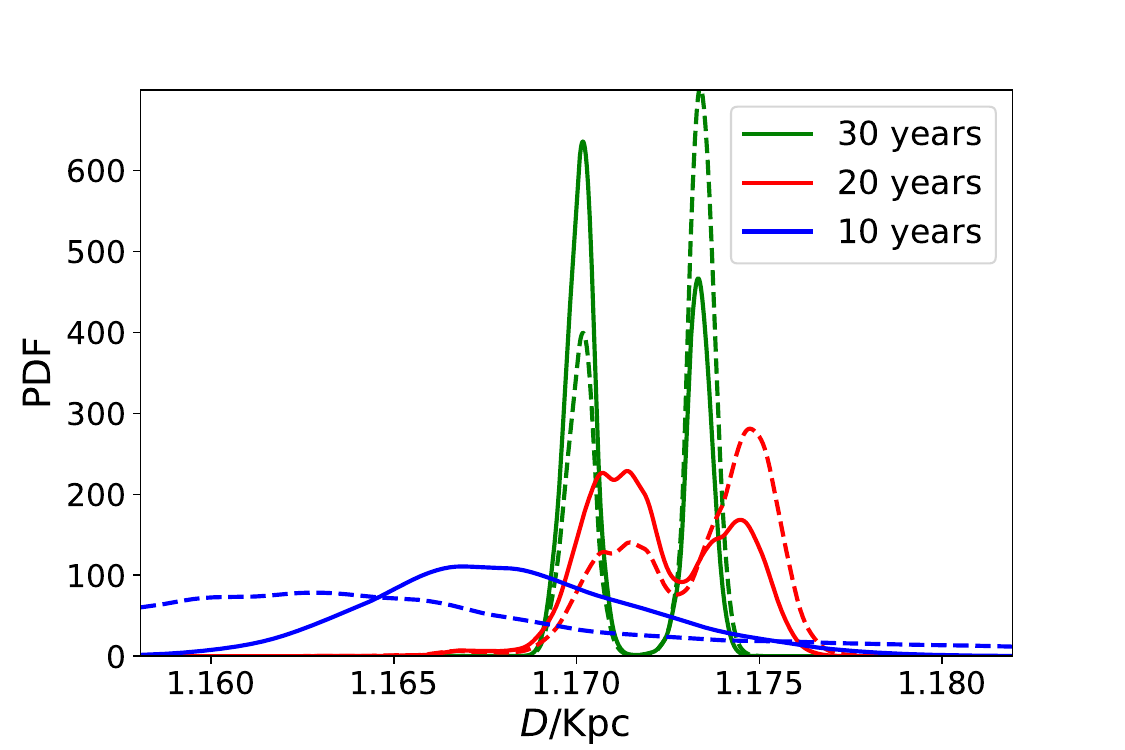}}
    \subfigure[J1911+1347]{\includegraphics[width=8cm]{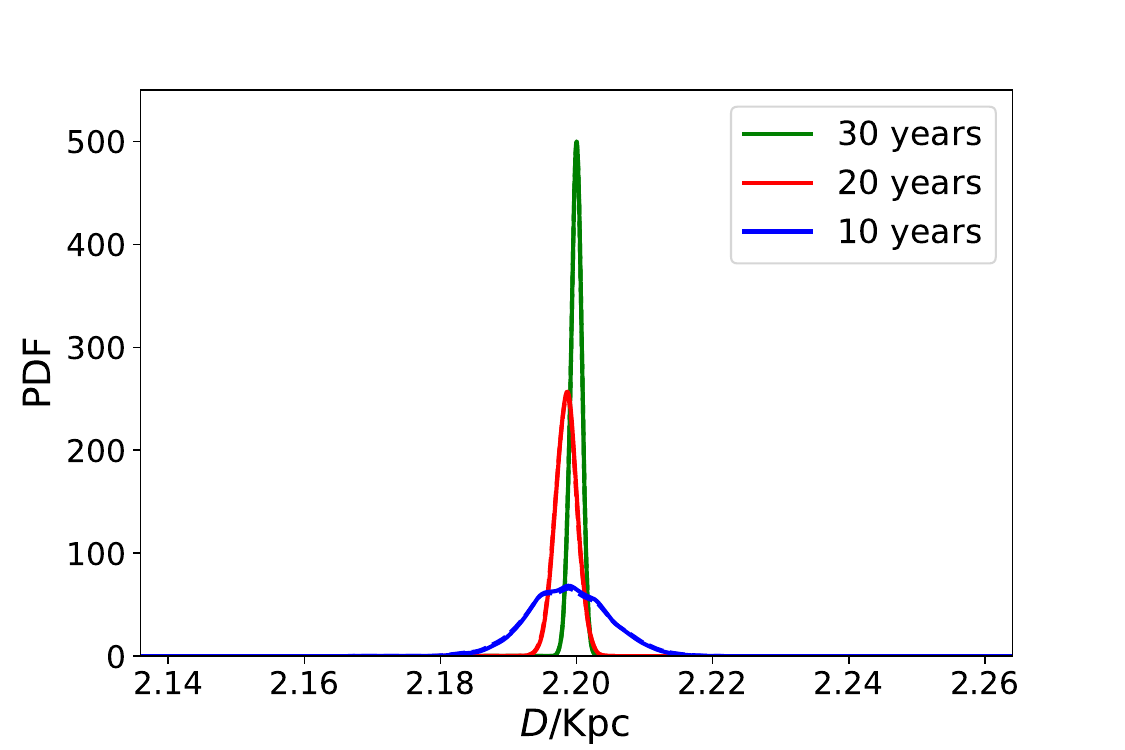}}
    \caption{The constraints on pulsar distance with the simulated SMBHB samples. The blue, red, green lines represent the results with $T=10,20, 30$ years, respectively. As a reference, the results without the priors of the timing parallax measurements are plotted in dashed lines. }
    \label{fig:multi_sample}
\end{figure*}

Similar to in the previous subsections, we use the MCMC method to quantify measurement precision of pulsar distances with nHz GWs from the selected SMBHBs (see Fig. \ref{fig:multi_sample} and Table \ref{tab:D_samples}). 
Assuming a short PTA observation span $T = 10$ years, the constraint uncertainties $\Delta D$ for pulsars with $D\sim 1$ kpc are found to be about $2-4$ pc. The variance in the distance measurement uncertainties of different pulsars is primarily attributable to (i) random orientations of sources relative to pulsars, and (ii) prior $\mathcal{\pi}(D)$ differences in parallax methods caused by pulsar distances and ecliptic latitude.  About half of the pulsars exhibits $\gtrsim20\%$ improvements in distance constraints. The largest improvement in distance measurement is found for J1911+1347 ($D=2.2$ kpc), with its distance uncertainty reduced by a factor of $\sim 3$ (from 14.87 pc to 5.55 pc), as shown in the right bottom panel of Fig. \ref{fig:multi_sample}. 

With a long PTA observation span  $T=30$ years, nHz GWs significantly improve the pulsar distance constraints, with $\Delta D < 0.4$ pc for 
18 pulsars (out of 20), and $\Delta D = 0.82$ pc even for J1911+1347 ( $\sim 15$ times better than the timing parallax measurement). 
It is worth note that there is still a double-peak structure in the distance constraint of J0751+1807 (left bottom panel of Fig. \ref{fig:multi_sample}). The reason is the number of
events that are effective in constraining the pulsar distance  $D$ strongly depends on 
the relative orientations between SMBHBs and pulsars, and under the set of samples simulated above, this number is too low to constrain J0751+1807's distance well. 
Excluding this specific case, GW observations achieve improvement factors of 2.5–15 in pulsar distance constraints across the remaining pulsars. 

As a comparison, we also show the pulsar distance constraints from PTA observations of nHz GWs alone without imposing  timing parallax priors  (Fig. \ref{fig:multi_sample}).  
The results are similar to the cases with timing parallax priors imposed, especially for $T\geq 20$ years, indicating that 
nHz GWs can potentially serve as an independent methodology for precision measurement of pulsar distances.

\begin{table}[ht]
    \centering
    \begin{tabular}{ccccc}\hline\hline
        \multirow{2}{*}{Pulsar}&\multirow{2}{*}{$D$(kpc)}&&$\Delta D$(pc)&\\
        && 10  years& 20 years & 30 years \\\hline
        J0030+0451&0.323&0.25(0.30)&0.12(0.21)&0.07(0.17)\\
        J0613-0200&0.99&3.32(3.48)&0.59(2.46)&0.29(2.01)\\
        J0751+1807&1.17&3.68(3.98)&2.07(2.81)&1.79(2.29)\\
        J1012+5307&1.07&3.64(5.45)&1.82(3.86)&0.38(3.15)\\
        J1022+1001&0.85&2.00(2.09)&0.64(1.48)&0.24(1.21)\\
        J1024-0719&0.98&2.67(3.01)&2.51(2.13)&0.28(1.74)\\
        J1455-3330&0.76&1.89(1.81)&0.37(1.28)&0.21(1.05)\\
        J1600-3053&1.39&5.14(5.77)&4.51(4.08)&0.37(3.33)\\
        J1640+2224&1.08&3.18(6.54)&0.76(4.63)&0.34(3.78)\\
        J1713+0747&1.136&3.06(5.06)&0.79(3.58)&0.36(2.92)\\
        J1730-2304&0.48&0.53(0.67)&0.23(0.47)&0.10(0.39)\\
        J1744-1134&0.388&0.38(0.46)&0.16(0.32)&0.08(0.26)\\
        J1751-2857&0.79&1.26(1.82)&0.39(1.29)&0.20(1.05)\\
        J1801-1417&1.0&1.87(2.97)&0.54(2.10)&0.28(1.72)\\
        J1804-2717&0.8&1.27(1.86)&0.39(1.32)&0.20(1.08)\\
        J1857+0943&1.11&2.26(5.00)&0.81(3.53)&0.34(2.89)\\
        J1909-3744&1.06&2.31(3.49)&0.56(2.47)&0.26(2.02)\\
        J1911+1347&2.2&5.76(21.36)&1.57(15.10)&0.82(12.33)\\
        J1918-0642&1.3&2.36(5.26)&0.89(3.72)&0.37(3.04)\\
        J2124-3358&0.47&0.52(0.71)&0.23(0.50)&0.12(0.41)\\

        \hline
    \end{tabular}
    \caption{The error bars of the pulsar distance measurements from PTA observations with different observation times. As a reference, the timing parallax measurements in Table \ref{tab:sigma_D} are also listed in brackets.}
    \label{tab:D_samples}
\end{table}

\section{Conclusion}
\label{sec:conclusions}

With strong evidence of stochastic nHz GWs recently reported by \cite{2023ApJ...951L...8A, 2023A&A...678A..50E, 2023ApJ...951L...6R, 2023RAA....23g5024X, 2025MNRAS.536.1467M},  
the detection of individual SMBHBs in the nHz band is also becoming foreseeable.  In general, the effect of a nHz GW source on pulsar timing can be decomposed as an earth term and a pulsar term,
where the pulsar term has usually be treated as a random noise because its phase $\Phi_{\rm p}\approx \omega D (1-\cos\theta)$ [Eq.~(\ref{eq_PhiP})] is subject to a large uncertainty if the pulsar distance $D$ is not well known. In this work, we have investigated how well the pulsar distances can be measured from PTA observations of nHz SMBHBs in the SKA era when the pulsar term is expected to be measured with a reasonable precision ($\Delta\Phi_{\rm P} \lesssim 1$, see Fig.~ \ref{fig:corner}). Since the pulsar term is a periodic function of $\Phi_{\rm p}$ with a period $2\pi$, the derived constraint on the pulsar distance $D$ is therefore periodic with a period $2\pi/[\omega(1-\cos\theta)]$.

One way to eliminate the periodicity in the pulsar distance constraint is incorporating extra information, e.g., the prior information from the timing parallax measurement.
To investigate the methodological feasibility, we perform MCMC simulations for a population of individual SMBHBs 
 with $\mathcal{M}_c=10^{10}\ M_\odot$, $d_L=5$ Gpc, $\omega_\mathrm{E}=10/30$ nHz. For pulsars with $D\lesssim1$ kpc, the priors from timing parallax measurements effectively suppress periodicity in distance constraints induced by individual SMBHBs. For a $T=30$ years observation period, these single-peaked golden events can improve timing parallax measurements by a factor of up to 4. For pulsars with $D\sim2$ kpc, the limited ability of timing parallax priors in eliminating periodicity leads to a low golden event fraction. However, these rare golden SMBHBs can  reduce pulsar distance uncertainties from $\sim10$ pc to $\sim1$ pc.

In addition to the prior information from the timing parallax measurement, the joint constraint from PTA observations of multiple SMBHBs is also useful in eliminating the false peaks in the pulsar distance constraint, since the period $2\pi/[\omega_i(1-\cos\theta_i)]$ in the pulsar distance constraint from each SMBHB differs.  As an example, we show in Fig. \ref{fig:multi} the joint constraint of the distance to pulsar J1911+1347
from 30 year PTA observations of different numbers of SMBHBs. 
To show the actual performance of this method, we then simulate a sample combing the SMBHB population models from \cite{2020ApJ...897...86C} and \cite{2023SCPMA..6620402B}. With this sample, 20 and 30 years of observations can constrain most pulsars' distances to 1 pc and 0.4 pc, respectively, which are much better than the accuracy of the timing parallax measurements.

In this work, we propose that future PTA observations of nHz SMBHBs is of great potential in precision measurement of pulsar distance measurement. 
Note that we have only considered a PTA of 20 pulsars as a proof of principle. In reality, more PTA pulsars  will further improve both the SMBHB parameters and the constraints of pulsar distances.



\section*{Acknowledgements}
The results in this paper have been derived using the following packages: \texttt{astropy}\cite{2022ApJ...935..167A}, \texttt{Bilby} \cite{2019ApJS..241...27A}, \texttt{corner} \cite{corner}, \texttt{dynesty} \cite{2020MNRAS.493.3132S}, \texttt{Jupyter} \cite{2016ppap.book...87K}, \texttt{numpy} \cite{2020Natur.585..357H}, \texttt{matplotlib} \cite{2007CSE.....9...90H}, \texttt{scipy} \cite{2020NatMe..17..261V}. This work made use of the Gravity Supercomputer at the Department of Astronomy, Shanghai Jiao Tong University.

\bibliographystyle{apsrev}
\bibliography{main}

\end{document}